\documentclass[sigconf, nonacm]{acmart}

\usepackage{graphicx}
\usepackage{balance}  

\usepackage{graphicx}
\usepackage{balance}  

\usepackage{color}
\usepackage{multirow}
\usepackage{subfigure}
\usepackage[export]{adjustbox}
\usepackage{enumitem}
\usepackage{subcaption}
\usepackage{hhline}
\usepackage{balance}
\usepackage{colortbl}
\usepackage{units}
\usepackage{marginnote}
\usepackage{pgfplotstable,array}
\usepackage{marginnote}
\usepackage{float}
\usepackage{pgfplots}
\usepackage{pgf,tikz}
\usepackage{mathtools}
\usepackage{xstring}
\usepackage{xcolor}
\usepackage{subcaption}
\usepackage{soul,color}

\usetikzlibrary{spy}
\usetikzlibrary{patterns}
\usepackage{xspace}
\usepackage{tabularx}  
\usepackage{graphicx}  

\newcommand{\stitle}[1]{\vspace{0.8ex}\noindent\textup{\textbf{#1}}}

\usepackage{algorithmicx}
\usepackage{algorithm}
\usepackage{algpseudocode}
\usepackage{scalerel}

\makeatletter
\newcommand{\currentfsize}{\f@size pt}
\makeatother

\newdimen\fsize

\newcommand{\eat}[1]{} 

\AtBeginDocument{%
  }
\newcommand\vldbdoi{XX.XX/XXX.XX}
\newcommand\vldbpages{XXX-XXX}
\newcommand\vldbvolume{14}
\newcommand\vldbissue{1}  
\newcommand\vldbyear{2020}
\newcommand\vldbauthors{\authors}
\newcommand\vldbtitle{\shorttitle} 
\newcommand\vldbavailabilityurl{URL_TO_YOUR_ARTIFACTS}
\newcommand\vldbpagestyle{plain} 

\begin{document}
\title{BEACON: A Benchmark for Efficient and Accurate Counting of Subgraphs [Experiment, Analysis \& Benchmark Paper]}
\author{Mohammad Matin Najafi$^{\dagger}$, Xianju Zhu$^\dagger$, Chrysanthi Kosyfaki$^\dagger$,  Laks V.S. Lakshmanan$^\S$, Reynold Cheng$^\dagger$}
\affiliation{
   \institution{$^\dagger$The
     University of Hong Kong, Hong Kong SAR, China}
     \country {}
   \institution{$^\S$The University of British Columbia, Vancouver, B.C., Canada\quad}
   \country {}
}
\email{mohammad,kosyfaki,ckcheng@cs.hku.hk;  zxj0302@connect.hku.hk;  laks@cs.ubc.ca}


\newcommand{\LL}[1]{\textcolor{blue}{#1}}
\newcommand{\laks}[1]{\textcolor{red}{[[From Laks: #1]]}}

\newcommand{\hlc}[2][yellow]{ {\sethlcolor{#1} \hl{#2}} }

\begin{abstract}

Subgraph counting—the task of determining the number of instances of a query pattern within a large graph—lies at the heart of many critical applications, from analyzing financial networks and transportation systems to understanding biological interactions. Despite decades of work yielding efficient algorithmic (AL) solutions and, more recently, machine learning (ML) approaches, a clear comparative understanding is elusive. This gap stems from the absence of a unified evaluation framework, standardized datasets, and accessible ground truths, all of which hinder systematic analysis and fair benchmarking.
To overcome these barriers, we introduce \textit{BEACON}: a comprehensive benchmark designed to rigorously evaluate both AL and ML-based subgraph counting methods. BEACON provides a standardized dataset with verified ground truths, an integrated evaluation environment, and a public leaderboard, enabling reproducible and transparent comparisons across diverse approaches.
Our extensive experiments reveal that while AL methods excel in efficiently counting subgraphs on very large graphs, they struggle with complex patterns (e.g., those exceeding six nodes). In contrast, ML methods are capable of handling larger patterns but demand massive graph data inputs and often yield suboptimal accuracy on small, dense graphs. These insights not only highlight the unique strengths and limitations of each approach but also pave the way for future advancements in subgraph counting techniques.
Overall, BEACON represents a significant step towards unifying and accelerating research in subgraph counting, encouraging innovative solutions and fostering a deeper understanding of the trade-offs between algorithmic and machine learning paradigms.

\end{abstract}

\maketitle

\pagestyle{\vldbpagestyle}
\begingroup\small\noindent\raggedright\textbf{PVLDB Reference Format:}\\
\vldbauthors. \vldbtitle. PVLDB, \vldbvolume(\vldbissue): \vldbpages, \vldbyear.\\
\href{https://doi.org/\vldbdoi}{doi:\vldbdoi}
\endgroup
\begingroup
\renewcommand\thefootnote{}\footnote{\noindent
This work is licensed under the Creative Commons BY-NC-ND 4.0 International License. Visit \url{https://creativecommons.org/licenses/by-nc-nd/4.0/} to view a copy of this license. For any use beyond those covered by this license, obtain permission by emailing \href{mailto:info@vldb.org}{info@vldb.org}. Copyright is held by the owner/author(s). Publication rights licensed to the VLDB Endowment. \\
\raggedright Proceedings of the VLDB Endowment, Vol. \vldbvolume, No. \vldbissue\ %
ISSN 2150-8097. \\
\href{https://doi.org/\vldbdoi}{doi:\vldbdoi} \\
}\addtocounter{footnote}{-1}\endgroup

\ifdefempty{\vldbavailabilityurl}{}{
\vspace{.3cm}
\begingroup\small\noindent\raggedright\textbf{PVLDB Artifact Availability:}\\
The source code, data, and/or other artifacts have been made available at \url{https://github.com/zxj0302/MLSC}.
\endgroup
}

\section{Introduction} \label{sec:intro}

Subgraph counting is a fundamental problem within graph analytics, where the objective is to enumerate and count all subgraphs in a large graph \(G = (V, E)\) that are isomorphic to a given query graph \(q = (V_q, E_q)\) \cite{foulds1995graph}. Formally, the goal is to compute \(C_G(q)\), or the number of occurrences of \(q\) within $G$. 


\begin{figure}[hbt]
    \centering
   \includegraphics[width=0.30\textwidth]{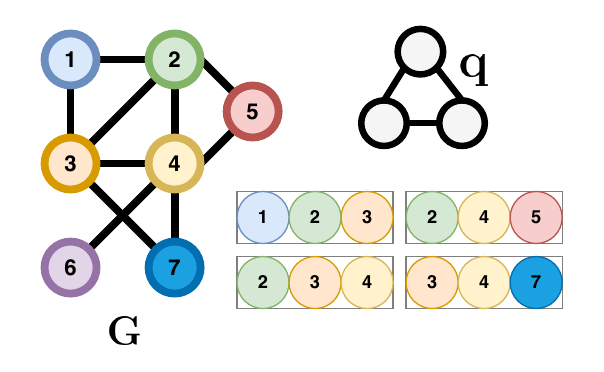}
     \vspace{-0.2in}
     \caption{An example input graph $G$, a query pattern $q$, and the $4$ corresponding instances of $q$ in $G$, yielding $\mathcal{N}_q(G) = 4$.}
     \vspace{-0.1in}
    \label{fig:intro}
  \end{figure}

Subgraph counting plays a vital role in many domains ~\cite{DBLP:conf/icalp/KaneMSS12,DBLP:journals/siamcomp/WilliamsW13,DBLP:conf/bigdataconf/AhmedWR16,DBLP:conf/www/SeshadhriT19,DBLP:conf/icdm/BordinoDGL08,DBLP:conf/icpp/SlotaM13,DBLP:journals/csur/RibeiroPSAS21,DBLP:journals/siamcomp/FockeR24,DBLP:conf/sigmod/BonifatiOTVY024,DBLP:journals/pvldb/FangLM22,DBLP:conf/cluster/RibeiroSL10,DBLP:journals/siamdm/FloderusKLL15,DBLP:conf/mlg/WackersreutherW10,DBLP:conf/icdm/KuramochiK01,DBLP:conf/edbt/PapapetrouIS11}. In network science, for example, identifying frequently-occurring patterns—such as cliques (fully connected subgraphs) or small recurring motifs—can help in deciphering community structures and interaction patterns within social networks \cite{chen2018mining}. In biological networks, counting subgraphs such as motifs is instrumental in revealing functional modules or biologically significant interactions \cite{wang2021learning,DBLP:journals/biodatamining/MrzicMBMCGL18,najafi23moser}. Furthermore, in financial and transaction networks, the detection and counting of specific subgraph patterns can expose anomalous transaction behaviors, which may signal fraudulent activities like money laundering \cite{DBLP:conf/edbt/KosyfakiMPT19}. These applications underscore the important role of subgraph counting for extracting deep insights about complex and interconnected systems.
Figure~\ref{fig:intro} illustrates a pattern \(q\), and its instances for graph $G$.



{\bf Challenges.} Due the NP-completeness of the underlying subgraph isomorphism, subgraph counting is computationally challenging~\cite{DBLP:journals/jacm/Ullmann76,DBLP:books/acm/23/Cook23a}.
This problem exacerbates when $G$ is huge, or when $q$ is complex. To tackle these issues, two classes of solutions were studied in the literature: Algorithmic (AL) and Machine Learning (ML)-based, as discussed below. 


{\bf 1. Algorithmic (AL) methods} include exact and approximate techniques. State-of-the-art exact methods leverage advanced combinatorial strategies and specialized hand-crafted formulas tailored to specific subgraph patterns (e.g., ESCAPE \cite{DBLP:conf/www/PinarSV17} and EVOKE \cite{DBLP:conf/wsdm/PashanasangiS20}), which can compute counts efficiently when such formulas exist. However, deriving these formulas is inherently challenging. Also, they usually  do not support larger or more complex subgraphs, and cannot scale to handle dense graphs due to the exponential nature of the NP-complete subgraph isomorphism ~\cite{DBLP:journals/jacm/Ullmann76,DBLP:books/acm/23/Cook23a}. 
Approximate AL methods (e.g., \cite{motivo}) aims to trade accuracy for efficiency. Their performance can degrades significantly as both $G$ and \(q\) become larger or denser. 



{\bf 2. Machine Learning (ML)-based approaches} have emerged as a promising alternative \cite{DBLP:conf/icml/PappW22,DBLP:conf/nips/QianRGN022,DBLP:conf/icassp/SandfelderVH21,DBLP:journals/pacmmod/SchwabeA24,DBLP:conf/aaai/SunHWW0Y24,DBLP:journals/pacmmod/ZhangSS023,DBLP:journals/tkde/ZhangBEZLYDYW23,zhao2021stars} to traditional algorithmic methods for subgraph counting. These methods use graph patterns to train a model (e.g., GNN~\cite{DBLP:conf/iclr/ZhaoJAS22}). The trained model can then produce subgraph counts quickly, even for large and complex 
graphs / subgraphs that cannot be supported by AL methods. However, there is a lack of a unified framework and curated datasets for ML-based subgraph counting, making standardized evaluation against AL strategies difficult. The performance of these models is also sensitive to variations in software environments, hardware configurations, and training procedures, raising reproducibility concerns. Moreover, we observe that in existing works in ML approaches, only few AL methods, which may not be SOTA, were used for comparisons. Without a comprehensive and systematic comparison, it is not clear {\it when} and {\it how} ML approaches fare better than AL methods. This makes it difficult to  realize the strengths and weaknesses of existing subgraph counting algorithms, or design better ones.

{\bf The BEACON benchmark.} To address the limitations and open questions of current subgraph counting approaches, we propose {\it BEACON}, or \underline{B}enchmark for \underline{E}fficient and \underline{A}ccuracy \underline{CO}u\underline{N}ting. This is the first unified framework that enables  comparisons among AL and ML-based methods across various types of graphs and query patterns. It also enhances reproducibility through standardized datasets, detailed protocols, and containerized environments. It can also be useful to the research community by elucidating the trade-offs among scalability, accuracy, and runtime that are made by the various methods. BEACON facilitates a deeper understanding of existing methodologies and supports the development of next-generation methods that effectively harness both algorithmic and machine learning techniques. Specifically, our contributions are: 


\noindent 
\textbf{(1) Insightful Experimental Findings:}

\noindent $\bullet$ \textbf{Comprehensive Evaluation Across Diverse Scenarios:}  
our benchmark evaluates a wide range of subgraph counting methods—including both algorithmic (AL) and machine learning (ML)-based approaches—under carefully controlled experimental conditions. By testing on a wide range of graphs in terms of size and density, our framework reveals interesting trends in accuracy, scalability, and robustness.

\noindent $\bullet$ \textbf{Detailed Trade-off Analysis:} With experiments conducted on a high-performance system, we measure preprocessing, training, and inference times. Our results uncover critical trade-offs: while lightweight GNN models achieve fast, sublinear inference times even on large graphs, methods such as PPGN \cite{ppgn}, though more accurate, incur significantly higher training costs. We further study how graph properties—such as density and degree distribution—affect metrics like Q-error and MAE across zero-shot, few-shot fine-tuning, and retraining scenarios.

\noindent $\bullet$ \textbf{Granular Insights into Performance Variability:}  The framework systematically distinguishes between local and global subgraph counts, as well as induced and non-induced patterns. This multidimensional analysis not only highlights the strengths and limitations of each method but also provides guidance for selecting or designing algorithms tailored to specific application needs.


\noindent 
\textbf{(2) Ease of Use and Reproducibility:}

\noindent $\bullet$ \textbf{Flexible Data Extraction with BEACON-Sampler:}  
To support our experiments, we integrate the \texttt{BEACON-Sampler}—a tool available on PyPI—which allows users to query and extract graphs from our extensive Oracle Dataset based on user-defined constraints like node count and average degree. This ensures that researchers can quickly generate customized benchmark datasets that suit their specific experimental scenarios.

\noindent $\bullet$ \textbf{Standardized, Open-Source Framework:}  
Our benchmark framework includes a suite of standardized datasets, detailed evaluation protocols, and an open-source code repository. This transparency guarantees that experiments can be easily reproduced, methods compared, and extended by the community. Researchers can leverage our clear separation of preprocessing, training, and inference phases to directly integrate new methods into our evaluation pipeline.

\noindent $\bullet$ \textbf{Community-Driven Public Leaderboard:}  
To foster ongoing research and collaboration, we provide a public leaderboard that tracks state-of-the-art performance on our benchmark. This shared resource not only motivates continuous improvement but also offers an immediate point of reference for comparing diverse subgraph counting algorithms.


To summarize, we establish a robust ecosystem for subgraph counting research—one that delivers in-depth experimental insights while providing an easy-to-use, reproducible platform that accelerates future developments in this challenging domain.

{\bf Roadmap.} The rest of the paper is organized as follows. In Section \ref{sec:notations-and-definitions}, we discuss the problem definitions and notations used in the paper. Section~\ref{sec:relwork} reviews related work on exact and ML-based subgraph counting approaches. Section~\ref{sec:fram} details our proposed framework. Section~\ref{sec:exps} presents experimental evaluations on diverse real-world networks using BEACON. Section~\ref{sec:concl} concludes.
\section{Notations and Definitions}
\label{sec:notations-and-definitions}

In this section, we introduce the primary notations and definitions used throughout the paper. We first provide a formal mathematical description of graphs and the concept of subgraph counting. Next, we distinguish between local and global subgraph counting approaches, subgraph-centric and network-centric methods, enumeration versus counting of subgraphs, and induced versus non-induced subgraph counting. Additionally, we define relevant error metrics, including Q-Error and Mean Absolute Error (MAE), and introduce the concept of the clustering coefficient.

\subsection{Graph Notation}
\label{sec:graph-notation}

A \emph{graph} $G$ is defined as an ordered pair $(V,E)$, where $V$ is a finite set of \emph{nodes} (or \emph{vertices}) and $E \subseteq \{\,(u,v) \mid u,v \in V,\, u \neq v \}$ is a set of \emph{edges}. 
Unless stated otherwise, we assume that $G$ is an undirected, simple graph: there are no self-loops (edges from a node to itself) and no parallel edges between the same pair of nodes.
We denote the number of nodes by $n = |V|$ and the number of edges by $m = |E|$.

\begin{definition}[Node and Edge]
\label{def:node-and-edge}
A \emph{node} $v \in V$ is one of the fundamental entities in the graph. An \emph{edge} $(u,v) \in E$ denotes a connection or relationship between the distinct nodes $u$ and $v$. 
\end{definition}

\subsection{Subgraph Counting}
\label{sec:subgraph-counting}

A \emph{subgraph} $H$ of $G$ is any graph whose node set and edge set are subsets of those in $G$. Formally, $H = (V_H, E_H)$ is a subgraph of $G = (V,E)$ if and only if $V_H \subseteq V$ and $E_H \subseteq E$. We are often interested in \emph{pattern-based} or \emph{motif} subgraphs that have a predetermined shape or size.

\begin{definition}[Subgraph Counting]
\label{def:subgraph-counting-def}
Let $F$ be a target subgraph (or pattern), and let $G$ be a graph. We define
\[
\mathcal{N}_F(G) \;=\; 
\bigl|\{\, H \subseteq G : H \cong F \}\bigr|,
\]
where $H \cong F$ denotes that $H$ is isomorphic to $F$. 
The \emph{subgraph counting} problem is to compute $\mathcal{N}_F(G)$. Two subgraph instances are considered distinct if they differ in at least one node or edge.
\end{definition}

\subsection{Local vs.\ Global Subgraph Counting}
\label{sec:local-global}

Subgraph counting can be performed at multiple levels of granularity:

\begin{enumerate}
    \item \emph{Local Subgraph Counting:} For each node $v \in V$, count the number of subgraphs that include $v$ and are isomorphic to $F$. Formally,
    \[
    \mathcal{N}_F(v; G) \;=\;
    \bigl|\{\, H \subseteq G : v \in V(H), \; H \cong F \}\bigr|.
    \]
    The result is a vector $\bigl(\mathcal{N}_F(v_1; G), \ldots, \mathcal{N}_F(v_n; G)\bigr)$, one entry per node.
    \item \emph{Global Subgraph Counting:} Count the total number of subgraphs isomorphic to $F$ in the entire graph $G$. This is simply
    \[
    \mathcal{N}_F(G) \;=\;
    \bigl|\{\, H \subseteq G : H \cong F \}\bigr|,
    \]
    yielding a single scalar value.
\end{enumerate}

\subsection{Subgraph-Centric vs.\ Network-Centric Methods}
\label{sec:subgraph-network-centric}

Algorithms for subgraph counting are often categorized based on their input and output:

\begin{enumerate}
    \item \emph{Subgraph-Centric Methods:} The algorithm is given a specific target subgraph (or pattern) $F$ as input and returns $\mathcal{N}_F(G)$. Examples include exact counting or estimation of a designated 4-node pattern with a particular connectivity structure.
    \item \emph{Network-Centric Methods:} The algorithm is instead given only the size $k$ of the subgraphs. The output is the number (or a vector of counts) of \emph{all} non-isomorphic subgraphs of size $k$ present in $G$. Formally, one might compute
    \[
    \bigl(\, \mathcal{N}_{F_1}(G), \; \mathcal{N}_{F_2}(G), \;\dots\,, \bigr),
    \]
    where each $F_i$ is a distinct unlabeled graph on $k$ nodes.
\end{enumerate}

\subsection{Subgraph Enumeration vs.\ Subgraph Counting}
\label{sec:enumeration-vs-counting}

Two common approaches to subgraph analysis differ in the level of detail they produce:

\begin{enumerate}
    \item \emph{Subgraph Enumeration:} Enumerate (i.e., explicitly list) \emph{all} subgraph instances isomorphic to $F$ (or of size $k$). The enumeration procedure might return a set 
    \[
    \{\, H_1, H_2, \dots, H_r \}\quad \text{where each }H_i \cong F,\;
    i=1,\dots,r.
    \]
    \item \emph{Subgraph Counting:} Merely compute the number of such subgraphs (i.e., $r$ in the above example), without enumerating the actual instances. Pure counting often yields computational savings compared to full enumeration.
\end{enumerate}

\subsection{Induced vs.\ Non-Induced Subgraph Counting}
\label{sec:induced-non-induced}

Let $S \subseteq V$ be a subset of nodes. The \emph{induced subgraph} of $G$ on $S$ is the graph $G[S] = (S, E_S)$ where
\[
E_S \;=\; \{\, (u,v) \in E : u \in S,\; v \in S \}.
\]
In other words, $G[S]$ includes all edges in $E$ whose endpoints lie entirely in $S$.

\begin{definition}[Induced Subgraph Counting]
\label{def:induced-subgraph}
\emph{Induced subgraph counting} restricts attention to subgraphs $H \subseteq G$ that are induced by their node sets. Formally, if $V(H) = S$, then $E(H)$ must be $E_S$ as in $G[S]$. 
\end{definition}

\begin{definition}[Non-Induced Subgraph Counting]
\label{def:non-induced-subgraph}
\emph{Non-induced subgraph counting} requires only that each subgraph $H \subseteq G$ matches a target pattern $F$ in node and edge configuration---but it need not include all edges among the chosen nodes. If $F$ has fewer edges than the complete graph on $V(F)$, the corresponding $H$ in $G$ may also omit those edges.
\end{definition}

\subsection{Q-Error vs.\ Mean Absolute Error (MAE)}
\label{sec:qerr-mae}

When estimating subgraph counts, it is crucial to quantify the discrepancy between an estimated count $\widehat{C}$ and the true count $C$. Two widely used metrics for subgraph count estimation are \emph{Q-error} and \emph{mean absolute error (MAE)}:

\begin{enumerate}
    \item \emph{Q-Error:} Given $C>0$ and $\widehat{C}>0$, the Q-error is
    \[
    \text{Q-error}(\widehat{C}, C) \;=\;
    \max \!\Bigl(\frac{\widehat{C}}{C}, \; \frac{C}{\widehat{C}}\Bigr).
    \]
    A small Q-error implies that $\widehat{C}$ and $C$ are close relative to each other, emphasizing multiplicative deviations.
    
    \item \emph{Mean Absolute Error (MAE):}
    \[
    \text{MAE} \;=\;
    \bigl|\widehat{C} - C\bigr|.
    \]
    MAE captures additive error, penalizing larger absolute discrepancies between the estimated and true values.
\end{enumerate}

\subsection{Clustering Coefficient}
\label{sec:clustering-coefficient}

The \emph{clustering coefficient} is another key metric that characterizes the extent to which nodes form tightly knit groups (often referred to as ``clusters'' or ``triangles'') in a graph.

\begin{definition}[Local Clustering Coefficient]
\label{def:local-cc}
For a node $v \in V$, let $\Gamma(v) = \{\,u \in V : (u,v) \in E\}$ be the set of neighbors of $v$. Suppose $|\Gamma(v)| = d_v$. The \emph{local clustering coefficient} of $v$, denoted $C(v)$, is defined as
\[
C(v) \;=\; \frac{\bigl|\{ \,(x,y) \in E : x,y \in \Gamma(v)\}\bigr|}{\binom{d_v}{2}},
\]
which is the fraction of possible edges among neighbors of $v$ that are actually present in $E$. 
\end{definition}

\begin{definition}[Global Clustering Coefficient]
\label{def:global-cc}
The \emph{global clustering coefficient} (also called the \emph{average clustering coefficient}) is often computed as the average of $C(v)$ over all nodes $v \in V$, i.e.,
\[
C_{\text{global}} \;=\; \frac{1}{n} \sum_{v \in V} C(v).
\]
\end{definition}
\section{Related Work}\label{sec:relwork}


The subgraph counting problem is well-studied in the literature \cite{DBLP:conf/nips/Chen0VB20,DBLP:conf/aaai/YouGYL21,DBLP:conf/kdd/YanZG0Z24,DBLP:conf/wsdm/PashanasangiS20,DBLP:journals/vldb/ZhaoYLZR23,DBLP:journals/pvldb/ZhangY0ZC20,DBLP:journals/pvldb/LiY24,DBLP:conf/dasfaa/ZhaoYHR23,DBLP:conf/aaai/YuL0023}. Given a graph $G$, the objective is to count occurrences of specific subgraphs within $G$. This problem can be challenging due to the size and complexity of the graph. As a result, there is a lot of interest in developing algorithms that use both exact and learning-based approaches to tackle these challenges. In this section, we summarize the most representative works in the area.

\stitle{Exact/Approximate solutions} Exact and approximate solutions have been extensively studied, particularly for their efficiency in counting subgraphs fast. These methods focus on reducing redundant computations and optimizing the search space to handle increasingly larger graphs.  Pinar et al., \cite{DBLP:conf/www/PinarSV17} propose an algorithic framework called ESCAPE to count arbitrary small graphs within a graph. They also extend their framework to produce exact counts for all 5-vertex subgraphs. ESCAPE is also capable of avoiding redundant enumeration and focuses only to a small set of patterns to compute all the 5-vertex subgraphs. 
Another  work that defines and studies the subgraph counting problem is \cite{DBLP:conf/wsdm/PashanasangiS20}. In the context of this work, the authors propose an algorithm called EVOKE. EVOKE introduces a novel approach by focusing on vertex orbits-specific configurations of vertices in subgraphs when counting 5-vertex subgraphs. By leveraging these orbits, EVOKE can more effectively capture structural patterns within graphs, reducing the computational burden compared to traditional exhaustive enumeration techniques.

Motivo \cite{DBLP:journals/pvldb/BressanLP19} proposes an approximate solution that applies color coding, a well-known technique for detecting subgraph patterns, to the subgraph counting problem. The main contribution in Motivo is its focus on reducing both the time and memory consumption associated with traditional color coding methods. The authors introduce an adaptive graphlet sampling strategy that targets rare or extreme cases of graphlets, allowing Motivo to efficiently count motifs in large-scale networks. This makes it particularly suited for real-world applications, where graphs can contain millions or billions of nodes and edges.

A recent work by Li and Yu \cite{DBLP:journals/pvldb/LiY24} approach subgraph counting from the perspective of local subgraph counting queries, which aim to count subgraphs in specific neighborhoods of a larger graph rather than globally. They propose a tree-decomposition-based algorithm that recursively breaks the graph into smaller subgraphs, applying symmetry-breaking techniques to reduce redundant counting. By incorporating symmetry-breaking rules, the algorithm can minimize the overhead of repeatedly counting identical substructures, resulting in a more efficient counting process.

\subsection{Machine Learning (ML)-Based Approaches}
The rise of graph machine learning has spurred significant interest in neural approaches for subgraph counting \cite{DBLP:conf/nips/FrascaBBM22,DBLP:conf/nips/AlsentzerFLZ20,DBLP:conf/aistats/TahmasebiLJ23}. Early works focused on understanding the fundamental capabilities of graph neural networks (GNNs), while recent innovations have introduced specialized architectures to overcome theoretical and practical limitations.

\textbf{GNN-based Models and Their Evolution}
The expressive power of standard GNNs for substructure counting was first formalized by \cite{DBLP:conf/nips/Chen0VB20}, who proved that message-passing networks (MPNNs) and 2-Weisfeiler-Lehman (2-WL) equivalent models cannot count connected induced substructures with $\geq 3$ nodes. This theoretical limitation motivated the development of Local Relational Pooling (LRP), which processes rooted egonets with permutation-invariant pooling to count local patterns. Though LRP demonstrated improved counting power, its computational complexity grows combinatorially with neighborhood size.

Subsequent works sought to enhance global pattern recognition while maintaining efficiency. \cite{DBLP:conf/kdd/LiuPHSJS20} introduced DIAMNet, which combines dynamic attention with external memory to count subgraph isomorphisms in linear time. By framing counting as a question-answering task between pattern and data graphs, DIAMNet achieved orders-of-magnitude speedups over backtracking algorithms while maintaining acceptable error margins. However, its sequence-based encoding struggles with complex topological relationships.

The quest for WL hierarchy-breaking expressivity led to novel architectures focusing on specific substructure classes. For cycle counting, \cite{DBLP:conf/iclr/HuangPMZ23} proposed I$^2$-GNNs, which augment subgraph MPNNs with dual-node identifiers to count 5- and 6-cycles—a task provably impossible for previous subgraph GNNs. Concurrently, \cite{DBLP:journals/tkde/XiaLL23a} derived countability conditions via hereditary tree-width analysis and introduced Layer Permutation Pooling (LPP), achieving 84\% error reduction on molecular datasets through localized graph decomposition and neural permutation schemes.

\textbf{Specialized Architectures for Scalability}
Recent advances prioritize computational efficiency without sacrificing counting power. \cite{DBLP:conf/kdd/YanZG0Z24} developed ESC-GNN, which replaces costly subgraph-wise GNN executions with precomputed structural embeddings encoding degree and distance distributions. This innovation preserves 3-WL equivalence while matching standard MPNNs' runtime complexity. For extreme-scale graphs, \cite{DBLP:conf/wsdm/Desco} proposed DeSCo—a canonical partitioning framework with heterogeneous message passing that reduces count prediction errors by 137$\times$ while enabling positional occurrence analysis.

\textbf{Challenges in ML Approaches}
Despite these advances, critical challenges remain. First, the field lacks standardized benchmarks for fair ML-vs-algorithm comparisons, as noted by \cite{DBLP:conf/aaai/YouGYL21}. Reproducibility is hampered by undocumented hyperparameter sensitivities and hardware dependencies—issues exacerbated in memory-intensive models like LRP \cite{DBLP:conf/nips/Chen0VB20}. Furthermore, most studies focus on isolated pattern families (e.g., cycles in \cite{DBLP:conf/iclr/HuangPMZ23}), making holistic performance assessments difficult. These gaps motivate our unified benchmarking framework in Section~\ref{sec:fram}.
\section{Proposed Framework}\label{sec:fram}
This section presents the details of our proposed benchmark framework, highlighting its modular structure, usage scenarios, and contribution opportunities. The framework is designed to facilitate the development, evaluation, and comparison of subgraph counting (SC) methods in a standardized and reproducible manner.

Our benchmark is composed of three primary modules: (1) \textbf{Data}, (2) \textbf{Environment}, and (3) \textbf{Testing}. These modules are designed to cover the full spectrum of tasks required for subgraph counting method development and evaluation, from data preparation to testing and comparison. Importantly, our evaluation framework is built to assess both traditional algorithmic techniques (AL) and modern machine learning–based (ML) approaches, ensuring a comprehensive and unified analysis across diverse methodologies.

\subsection{Framework Overview}
\begin{figure*}[t!]
  \centering
  \subfigure[Framework]{
   \label{fig:framework}
    \includegraphics[width=0.27\textwidth]{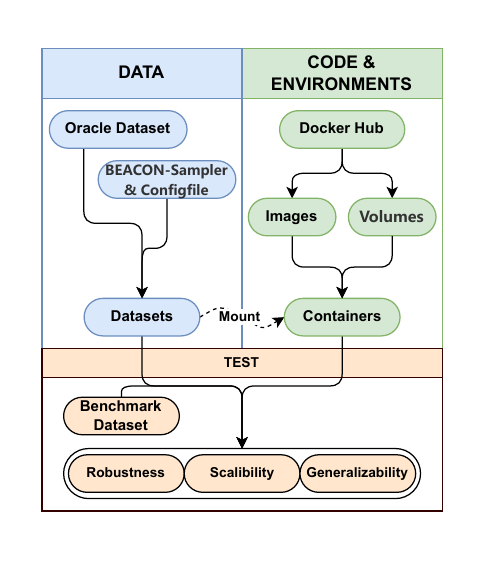}
    }\!\!
  \subfigure[Scenarios]{
   \label{fig:scenarios}
    \includegraphics[width=0.56\textwidth]{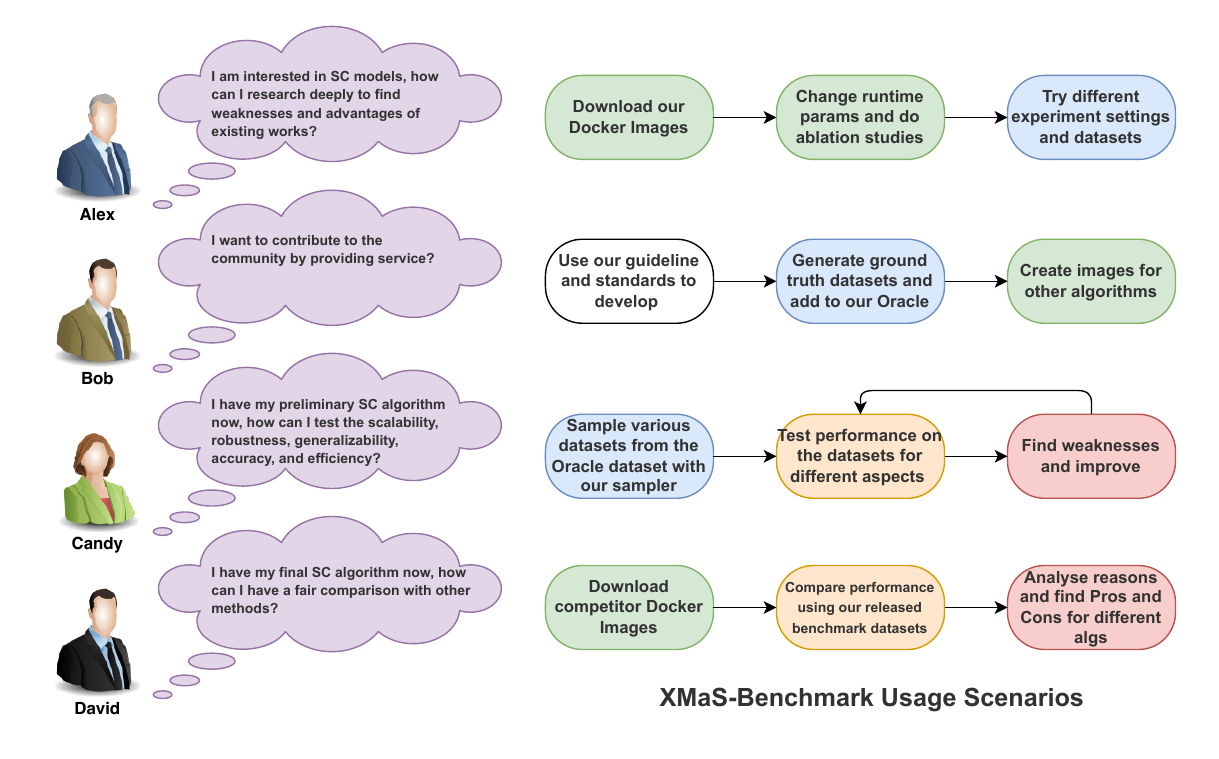}
    }
    \vspace{-0.1in}
  \caption{(a) The framework of the proposed benchmark, and (b) different usage scenarios.}
  \vspace{-0.2in}
  \label{fig:frameworkAndSecnario}
\end{figure*}

The core architecture of our framework is illustrated in Figure~\ref{fig:framework}. Each module serves a specific purpose within the framework:

\textbf{Data Module}
The data module is responsible for providing datasets required for training, validation, and testing subgraph counting methods. The module includes an interface, referred to as the BEACON-Sampler, which allows users to sample datasets based on specific user-defined constraints such as subgraph size, graph density, or the type of graph (e.g., directed or undirected). The sampler ensures that datasets are selected in a way that aligns with the user's research objectives, depending on whether they are testing scalability, accuracy, or generalizability. 

\textbf{Environment Module}
This module provides a controlled and reproducible environment for method development and evaluation. It leverages Docker containers to encapsulate the entire software stack, including dependencies, libraries, and code. By using containers, we ensure that methods can be tested and compared in identical environments, eliminating inconsistencies due to variations in system configurations or software versions. This module also includes a Docker Hub page\footnote{\url{https://hub.docker.com/repository/docker/zhuxiangju/benchmark_subgraphcounting/general}} where users can upload and share their own Docker images, ensuring that new methods adhere to the standards of reproducibility.

\textbf{Testing Module}
The testing module is designed to evaluate the performance of subgraph counting methods across several dimensions, including \textit{robustness}, \textit{scalability \& efficiency}, \textit{accuracy}, and \textit{generalizability}. 
It uses a set of pre-defined benchmark datasets, which are representative of various real-world scenarios, to assess how well methods perform under different conditions. This module also includes a leaderboard where users can compare their method's performance with state-of-the-art methods in a fair and transparent manner. 

\subsection{Usage Scenarios}

Our benchmark is designed to support a wide range of usage scenarios, allowing researchers and developers to utilize the framework for various purposes depending on their goals. Below, we outline four key usage scenarios, each addressing a different stage in the development lifecycle of a particular method for subgraph counting. 

\textbf{Dataset Usage Only}
In this scenario, users are primarily interested in utilizing the existing ground truth datasets provided by the benchmark for their own downstream applications, without necessarily engaging in method development. For example, a researcher might use the benchmark datasets to validate a hypothesis or compare two existing solutions. The datasets come pre-labeled with ground truth subgraph counts, allowing users to focus on their specific application without the need for further data processing.

\textbf{Explore the Literature}
This scenario targets users who are new to the field of subgraph counting and want to explore the existing literature and techniques. Users can download the datasets and images, modify parameters, and experiment with different settings to gain a deeper understanding of how the methods work. Additionally, users can access the results of our benchmark tests and compare them to existing methods. This scenario is ideal for researchers who want to familiarize themselves with the field before developing their own methods.

\textbf{Method Development}
In this scenario, users are interested in developing new subgraph counting methods. They can use the benchmark to perform a literature review, explore existing results, and use our datasets for preliminary testing. The BEACON-Sampler allows users to generate datasets tailored to their method's needs, providing a flexible platform for experimentation. Once the method is developed, users can use the framework's leaderboards to compare their method's performance against state-of-the-art methods.

\textbf{Method Benchmarking}
This scenario is designed for users who already have a fully developed and nearly publishable method. They can use the benchmark to rigorously evaluate their method's strengths and weaknesses in a fair and controlled environment. By comparing their method against others in the benchmark, users can identify areas for improvement and gain recognition on the benchmark's leaderboard. This scenario is particularly suited for researchers looking to publish their work in competitive venues.

Figure~\ref{fig:scenarios} illustrates these scenarios, showing how different users (e.g., Alex, Bob, Candy, and David) interact with the benchmark framework to achieve their respective goals. Each user represents a different stage in the method development lifecycle, from initial exploration to final benchmarking.

\subsection{Contribution Scenarios}

In addition to usage scenarios, our benchmark encourages active contributions from the research community. Below, we outline several ways in which users can contribute to the benchmark, ensuring that it continues to grow and improve over time.

\textbf{Add New Datasets}
Users can contribute new datasets to our Oracle dataset by either finding real-world datasets or generating synthetic ones. These datasets should include ground truth subgraph counts, which will be reviewed by our team before being added to the benchmark. This contribution helps expand the diversity of datasets in the benchmark, making it more representative of real-world applications.

\textbf{Improve Existing Datasets}
Users can improve existing datasets by calculating ground truth subgraph counts for larger subgraphs or more complex graphs. These contributions help enhance the accuracy and depth of the benchmark, making it more useful for testing advanced methods. 

\textbf{Pipeline Contribution}
Our benchmark pipeline is fully open-source, allowing users to contribute improvements to the codebase. For example, users can optimize the pipeline for better performance, add new features, or fix bugs. These contributions help ensure that the benchmark remains up-to-date with the latest advancements in software development.

\textbf{Docker Hub Contributions}
Reproducibility is a key focus of our benchmark, and we encourage users to publish their methods in Docker containers. By uploading their Docker files to our Docker Hub page, users can ensure that their methods can be easily tested and compared by others in the community. This contribution helps maintain a standard for reproducibility across the field.

\subsection{Modular Design Benefits}
The modular design of our benchmark offers several key advantages. First, it provides flexibility by allowing users to interact with the framework at various stages of the method development lifecycle, from dataset curation and preparation to full-scale benchmarking. Second, reproducibility is ensured through the use of Docker containers, which guarantee that all methods—whether traditional methods or machine learning–based approaches—are tested in identical environments, thereby eliminating variability due to system configuration differences. Third, our framework scales gracefully with the complexity of the task: the BEACON-Sampler supports the generation of datasets in a wide range of sizes and complexities, making it suitable for evaluating methods designed for large-scale graphs. Additionally, the open-source and community-driven nature of the benchmark encourages ongoing contributions and improvements. Finally, our design provides a dedicated avenue for application case studies, addressing a critical need in the field where subgraph analysis can considerably benefit from real-world validation and use-case enhancement.


\section{Experimental Evaluation}\label{sec:exps}
In this section, we showcase how our benchmark framework and curated datasets can be used to extract meaningful insights from the literature. Using our comprehensive platform and datasets, we easily compare diverse subgraph counting methodologies and reveal trends and nuances that might otherwise remain hidden. This approach not only promotes transparency and reproducibility but also underscores the essential role of a unified benchmark in advancing research and collaboration within the subgraph counting community. All methods were implemented in Python
and the experiments were conducted on a high-performance system equipped with an Intel Core i9-14900KF processor, 256 GB of RAM, and dual NVIDIA RTX 4070 SUPER GPUs. The source code of the paper is publicly available \footnote{https://github.com/zxj0302/MLSC}.






\subsection{Dataset}

In this section we discuss our benchmark's Data Module, which consists of three main parts: the Oracle Dataset, the BEACON-Sampler, and the Benchmark Dataset. We discuss each in detail in what follows.

\textbf{Oracle Dataset}

A key motivation for our Oracle Dataset is the absence of a large-scale dataset with ground truth subgraph counts in the community. To address this gap, we collected all graphs from the TUDataset \cite{Morris2020TUDataset} alongside graphs from the OGB dataset \cite{hu2020ogb}, which encompasses a variety of domains such as bioinformatics, social networks, and computer vision. For each graph, we computed the ground truth counts for all subgraphs with up to five nodes, considering both local and global frequencies as well as induced and non-induced configurations. We have made this extensive collection publicly available, and we refer to it as the \textit{Oracle Dataset}.



Our Oracle Dataset contains a total of 26,435 graphs drawn from multiple domains. In addition to the standard ground truth subgraph counts, we have augmented the dataset with the ID-constrained ground truth introduced by the DeSCo algorithm \cite{DBLP:conf/wsdm/Desco}. To further characterize each graph, we computed several graph-level features including diameter, density, and clustering coefficient.

\begin{figure}[h]
    \centering
    \includegraphics[width=0.65\columnwidth,height=5.5cm]{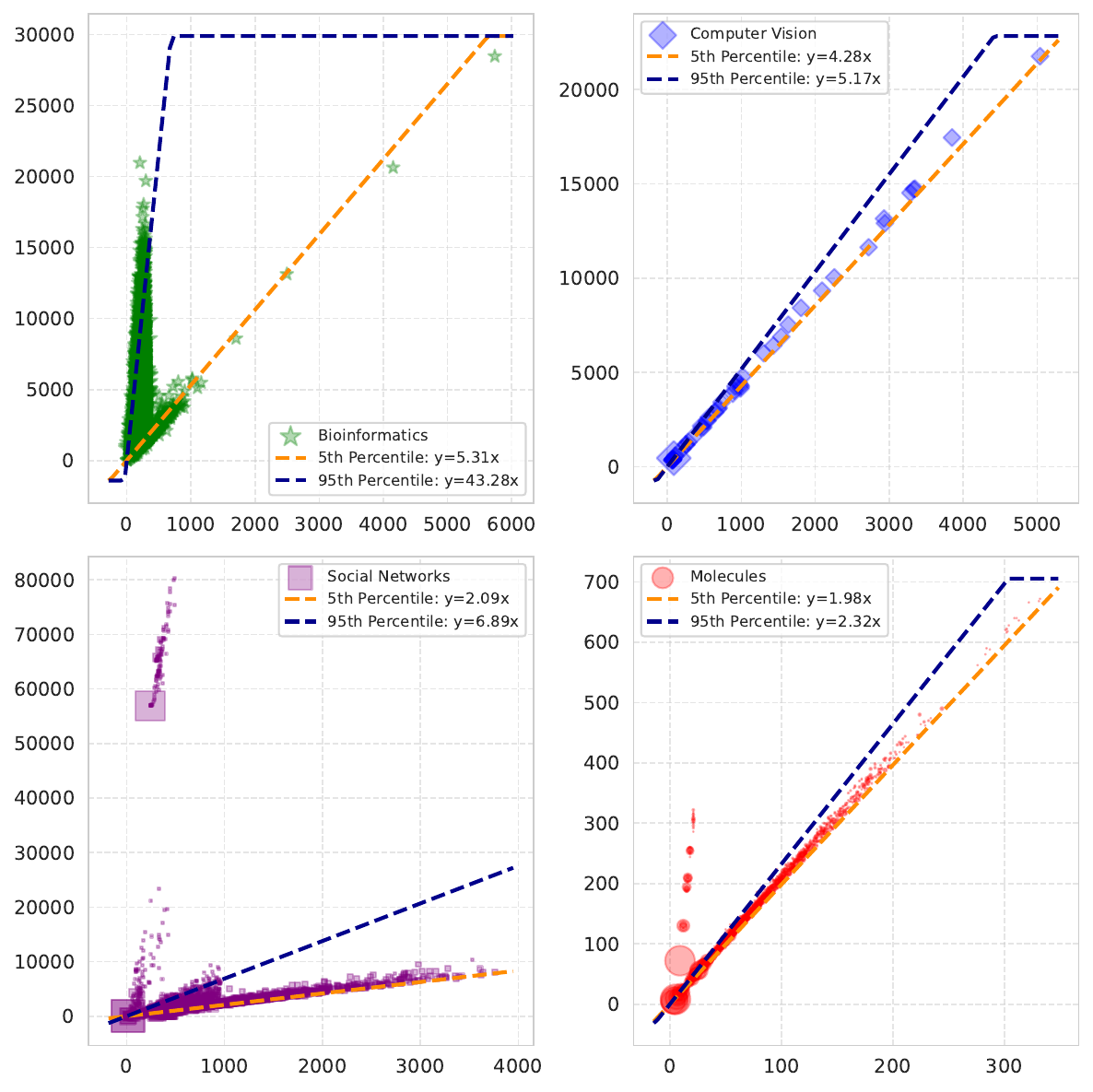}
    \vspace{-0.2in}
    \caption{Oracle Dataset's Domain and Density Distribution.}
    \label{fig:dataset_dist}
    \vspace{-0.15in}
\end{figure}

Figure~\ref{fig:dataset_dist} illustrates the distribution of node and edge counts across domains, showing the 5th and 95th percentile trends. The slope of these percentile lines—reflecting the ratio $\frac{|E|}{|V|}$—varies significantly between domains. Bioinformatics graphs, for example, tend to have higher densities as indicated by a steeper slope, whereas molecular networks generally exhibit lower densities. These observations emphasize the structural diversity present in the Oracle Dataset, making it a valuable resource for evaluating subgraph counting methods across a broad spectrum of network types.

Given that the Oracle Dataset comprises around 24,000 graphs with a wide range of characteristics, it is often impractical to utilize the full dataset in every research application. In many cases, researchers require a carefully selected subset that meets specific experimental criteria. For example, a study might need a sample of 10 social networks with a density greater than 3 and a node count between 20,000 and 50,000. To meet these tailored needs, we have developed the BEACON-Sampler, a tool that enables users to efficiently downsample the Oracle Dataset based on their given constraints.

\textbf{BEACON-Sampler}

The BEACON-Sampler is a versatile tool for extracting graphs from a database based on specific structural and numerical criteria. Publicly available on PyPI\footnote{https://pypi.org/project/rwdq/} under the name \texttt{rwdq}, this tool enables researchers to tailor their dataset selection through a JSON configuration file. Users can specify constraints such as minimum and maximum node counts, average degree thresholds, and other key graph properties.

This configuration-driven approach permits precise filtering of graphs to match particular experimental needs. For instance, researchers can define numerical thresholds that ensure the extraction of graphs with targeted density or degree distributions, which is especially important in tasks like subgraph counting where algorithm performance is sensitive to underlying graph characteristics.

Furthermore, the BEACON-Sampler is designed to be compatible not only with our Oracle Dataset but also with any dataset that conforms to our established format. This flexibility supports a wide range of applications, from systematic stress testing to targeted experimentation on specific graph structures, thereby facilitating optimized and reproducible research workflows.

Building on the foundations provided by the Oracle Dataset and the flexibility offered by the BEACON-Sampler, we now present the Benchmark Dataset. The idea behind this benchmark is to curate different subsets with varying sizes, densities, and characteristics. Each set is designed to test different aspects of subgraph counting methods—from scalability and efficiency to robustness—providing a straightforward platform for practical evaluation.

\textbf{Benchmark Dataset}

Using the BEACON-Sampler, we have curated a \textit{Benchmark Dataset} to evaluate subgraph counting algorithms from multiple angles, including scalability, accuracy, and sensitivity. The dataset is organized into sets with distinct constraints, as detailed in Tables~\ref{tab:dataset_chars_small} and \ref{tab:dataset_chars_large}.

Sets 1 through 6 (see Table~\ref{tab:dataset_chars_small}) are tailored to assess performance on smaller graphs with varying densities. These sets are split into training, validation, and test subsets in a 4:1:1 ratio. In Sets 1, 3, and 5, a maximum degree of 100 is enforced to obtain graphs with progressively increasing density. Dense graphs tend to exhibit features like cliques—challenging patterns that often remain absent in sparser datasets such as MUTAG. In contrast, Sets 2, 4, and 6 impose no maximum degree limit, better reflecting real-world networks where highly connected central nodes are common.

This division into many small graphs not only allows for easier parallel processing (as seen with algorithms like ESCAPE~\cite{escape}) but also reduces the memory footprint, since each graph can be handled independently.

\begin{table}[ht]
    \small
    \centering
    \begin{tabular}{|c|c|c|c|c|c|c|}
        \hline
        \textbf{Constraints} & \textbf{Set\_1} & \textbf{Set\_2} & \textbf{Set\_3} & \textbf{Set\_4} & \textbf{Set\_5} & \textbf{Set\_6} \\
        \hline
        \#nodes        & 0-500   & 0-500   & 0-500   & 0-500   & 0-500   & 0-500    \\
        \hline
        degree\_avg    & 0-1.5   & 0-1.5   & 1.5-5   & 1.5-5   & 5-10    & 5-10     \\
        \hline
        degree\_max    & 100     & None    & 100     & None    & 100     & None     \\
        \hline
        \#graphs       & 1200    & 1200    & 1200    & 1200    & 1200    & 1200     \\
        \hline
    \end{tabular}
    \caption{Small sets (Sets 1 to 6).}
    \label{tab:dataset_chars_small}
\end{table}

For scalability experiments, Sets 7 through 10 (see Table~\ref{tab:dataset_chars_large}) comprise larger graphs, divided into training, validation, and test subsets in a 4:1:4 ratio. These sets are designed to monitor how inference time and accuracy evolve with increasing graph size—helping to identify whether an algorithm scales efficiently or encounters resource constraints with larger graphs.

Overall, this dual approach—combining many small graphs with fewer, larger ones—ensures that our Benchmark Dataset provides a comprehensive framework to test the computational and memory challenges inherent in subgraph counting.

\begin{table}[ht]
    \small
    \centering
    \begin{tabular}{|c|c|c|c|c|}
        \hline
        \textbf{Constraints} & \textbf{Set\_7} & \textbf{Set\_8} & \textbf{Set\_9} & \textbf{Set\_10} \\
        \hline
        \#nodes        & 100-5000 & 100-5000 & 100-5000 & 100-5000 \\
        \hline
        \#graphs       & 200      & 500      & 1000     & 2000     \\
        \hline
    \end{tabular}
    \caption{Large sets (Sets 7 to 10).}
    \label{tab:dataset_chars_large}
\end{table}

\subsection{Setup}

In this section, we detail the experimental setup used to evaluate the performance of different methods using our benchmark. A structured and transparent setup is essential to ensure reproducibility and to provide a clear understanding of the conditions under which our results were obtained. This includes the patterns used for testing, the competitors, and the training procedures we employed. Each component of the setup has been carefully chosen to reflect real-world scenarios and to provide comprehensive insights into the method’s capabilities under different conditions. First, we describe the graph patterns used in our experiments and the rationale behind selecting them, followed by an overview of the competing methods we tested against. Finally, we present the details of our training process, including the parameters used and the learning strategies explored.

\textbf{Tested Patterns}
\begin{figure}[h]
      \centering
      \includegraphics[width=\columnwidth]{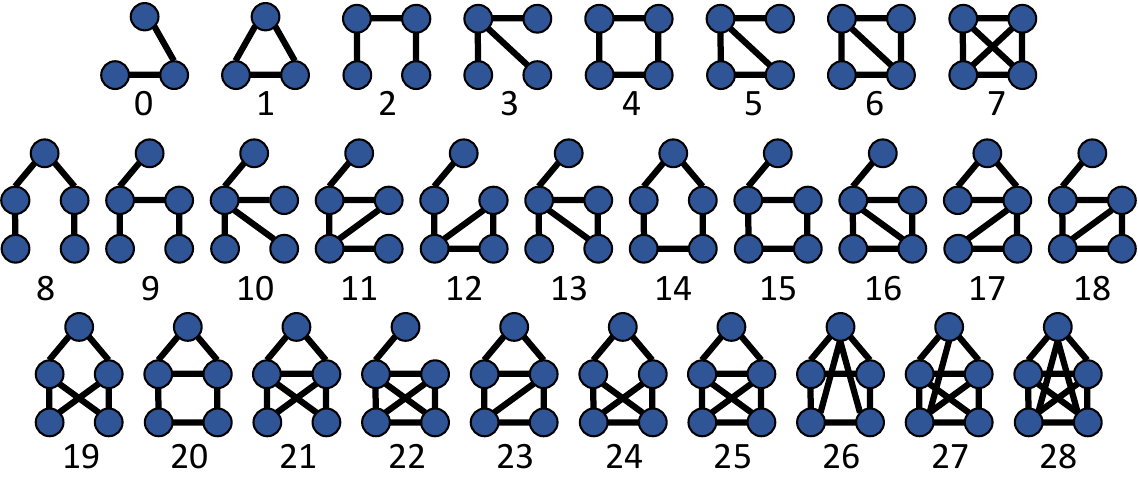}
      \vspace{-0.2in}
      \caption{Tested patterns (2 to 5 nodes).}
      \vspace{-0.17in}
      \label{fig:patterns}
\end{figure}

To assess the performance of each methods, we used a diverse set of patterns, ranging in size from 2 to 5 nodes. These sizes were selected because they are supported by all competing algorithms, including exact methods, and have been shown to provide valuable insights in practical applications \cite{DBLP:journals/csur/RibeiroPSAS21}. Moreover, larger patterns are not supported by some existing algorithms (e.g., ESCAPE), and calculating ground truth for these patterns is currently infeasible on large graphs. Including such patterns would significantly expand the complexity of our experiments due to the vast number of possible patterns. As a result, we opted to exclude them for now. However, we welcome any contributions from the community that could help extend our dataset.

The specific patterns used in our experiments, along with their naming conventions, are shown in Figure~\ref{fig:patterns}. For clarity, we will refer to these patterns as ``target'' moving forward. For instance, target $1$ refers to the triangle pattern.

\textbf{Competitors}
As shown in Table~\ref{tab:algorithms}, we evaluated a broad range of subgraph counting methods using our benchmark, classifying them into three categories: exact, approximate, and ML-based. Exact algorithms, such as ESCAPE and EVOKE, can compute accurate subgraph counts but are constrained by computational limitations, particularly for larger subgraph sizes (denoted by the parameter $k$). Approximate algorithms, like MOTIVO, prioritize speed over accuracy, making them suitable for larger graphs. Finally, ML-based algorithms, including GNN-based methods like PPGN and IDGNN, harness machine learning to offer scalable and adaptable solutions.

Table~\ref{tab:algorithms} categorizes these methods based on year, k-restriction (``-'' indicates no theoretical restriction), type, induction type (I.T.: non-induced (N), induced (I), or both (B)), counting process (C.P.: local (L) or global (G) subgraph counting), centricity type (C.T.: subgraph-centric (S) or network-centric (N)), and code repositories. Induced counts can be converted to non-induced and vice versa under the conditions in \cite{DBLP:conf/www/PinarSV17}. Global counts can be obtained by summing local counts and dividing by the redundancy factor.

\begin{table}[ht]
    \centering
    \resizebox{\linewidth}{!}{ 
        \begin{tabular}{|c|c|c|c|c|c|c|c|}
        \hline
        \textbf{Algorithm} & \textbf{Year} & \textbf{k} & \textbf{Type} & \textbf{I.T.} & \textbf{C.P.} & \textbf{C.T.} & \textbf{Code} \\
        \hline
        ESCAPE \cite{DBLP:conf/www/PinarSV17} & 2017 & 5 & Exact & N & G & N & \cite{escape} \\
        \hline
        EVOKE \cite{DBLP:conf/wsdm/PashanasangiS20} & 2020 & 5 & Exact & N & L & N & \cite{evoke} \\
        \hline
        MOTIVO \cite{DBLP:journals/pvldb/BressanLP19} & 2019 & - & Approx. & I & G & N & \cite{motivo} \\
        \hline
        GNN \cite{DBLP:conf/iclr/XuHLJ19} & 2019 & - & ML-based & B & L & S & \cite{gnn} \\
        \hline
        GNNAK \cite{DBLP:conf/iclr/ZhaoJAS22} & 2022 & - & ML-based & B & L & S & \cite{gnnak} \\
        \hline
        IDGNN \cite{DBLP:conf/iclr/ZhaoJAS22} & 2022 & - & ML-based & B & L & S & \cite{idgnn} \\
        \hline
        PPGN \cite{DBLP:conf/iclr/ZhaoJAS22} & 2021 & - & ML-based & B & L & S & \cite{ppgn} \\
        \hline
        I2GNN \cite{DBLP:conf/iclr/HuangPMZ23} & 2022 & - & ML-based & B & L & S & \cite{i2gnn} \\
        \hline
        ESC-GNN \cite{DBLP:conf/kdd/YanZG0Z24} & 2024 & - & ML-based & B & L & S & \cite{esc} \\
        \hline
        DeSCo \cite{DBLP:conf/wsdm/Desco} & 2024 & - & ML-based & B & L & S & \cite{descocode} \\
        \hline
        \end{tabular}
    }
    \caption{Overview of methods.}
    \label{tab:algorithms}
\end{table}

\textbf{Training Process}

The selection of hyperparameters is critical for achieving optimal performance and ensuring a fair comparison across all algorithms. We chose the hyperparameters based on both empirical evaluations and guidelines from the literature (suggestions from the authors), aiming to balance accuracy, training time, and generalization. For each method, we conducted a hyperparameter search within reasonable ranges, taking into account factors such as the number of layers, learning rate, and batch size. Further details about the specific parameters used for each algorithm are available on our website.

In a typical subgraph counting task on a static graph, the counting operation only needs to be performed once. Training a model on the same graph beforehand is impractical for two reasons. First, the training process is time-consuming. Second, effective training requires access to at least a portion of the ground truth, and if that is available, then the use of an ML model for prediction becomes unnecessary. Instead, an ideal ML approach is one that is trained in advance using established datasets, and can then accurately predict subgraph counts on completely unseen graphs. 

To assess this capability, we adopt a zero-shot evaluation where models—pre-trained on separate data—are directly applied to our Benchmark Dataset. We also consider scenarios where a model is given a limited look at the new data: one where a pre-trained model is fine-tuned for a few epochs (few-shot training), and another where a model is trained from scratch in a few-shot setting to test its unprimed adaptability. Finally, we evaluate each algorithm under a full training setup to observe its peak performance, even though such a scenario is less practical for static graphs. 
In what follows, we discuss these evaluation scenarios in detail.

We utilize the dataset from \cite{DBLP:conf/iclr/HuangPMZ23} and pretrain our models on this dataset. Our evaluation strategy is divided into four scenarios:

\textbf{Zero-Shot Evaluation}: In this phase, we test the pre-trained model directly on the Benchmark Dataset without any additional fine-tuning. This approach evaluates the model's ability to generalize to completely unseen graphs and confirms that it has learned robust subgraph counting patterns rather than simply memorizing its training data.

\textbf{Few-Shot Learning (Fine-Tuning Pretrained Models)}: Here, we fine-tune a pre-trained model using a small portion of the new dataset—up to 10\% of the full training epochs. This limited fine-tuning examines how effectively the model leverages its prior knowledge to quickly adapt to new data and improve its accuracy.

\textbf{Few-Shot Learning (Training from Scratch)}: In this scenario, we train models from scratch with only a few training epochs. By comparing these results with the fine-tuned models, we assess whether pre-training offers a substantial advantage in learning generalized subgraph counting patterns on unseen graphs.

\textbf{Full Training}: Lastly, models are fully trained from scratch to determine their best achievable performance. Although this setup is less practical for static graphs—where the counting task is performed only once—it provides a valuable baseline for comparing the peak performance of different models.

\begin{figure*}[t!]
  \centering
  \subfigure[Zero-Shot]{
   \label{fig:zero_shot}
    \includegraphics[width=0.31\textwidth]{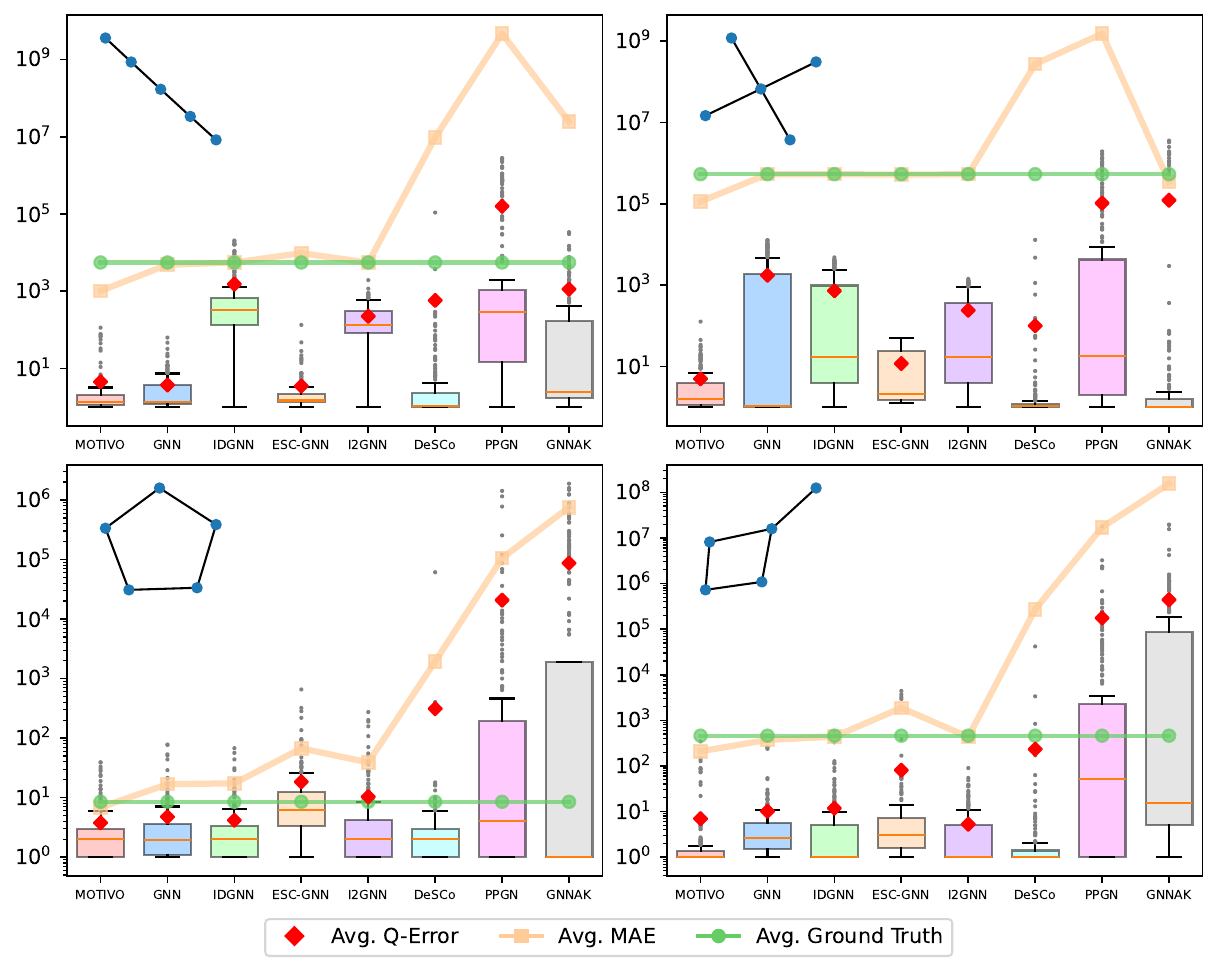}
    }\!\!
  \subfigure[Few-Shot Finetune]{
   \label{fig:few_shot}
    \includegraphics[width=0.31\textwidth]{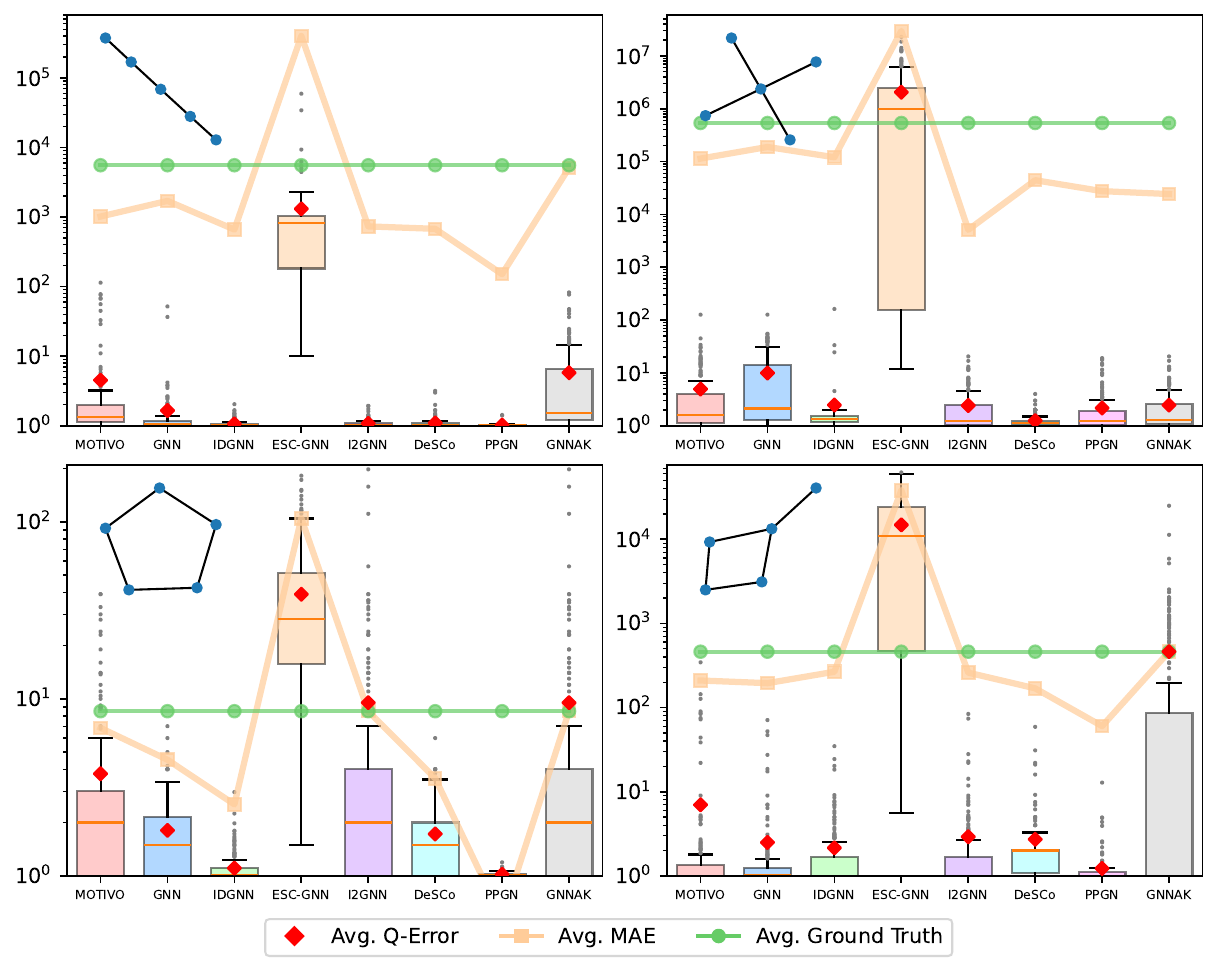}
    }
  \subfigure[Few-Shot Retrain]{
   \label{fig:retrain}
    \includegraphics[width=0.31\textwidth]{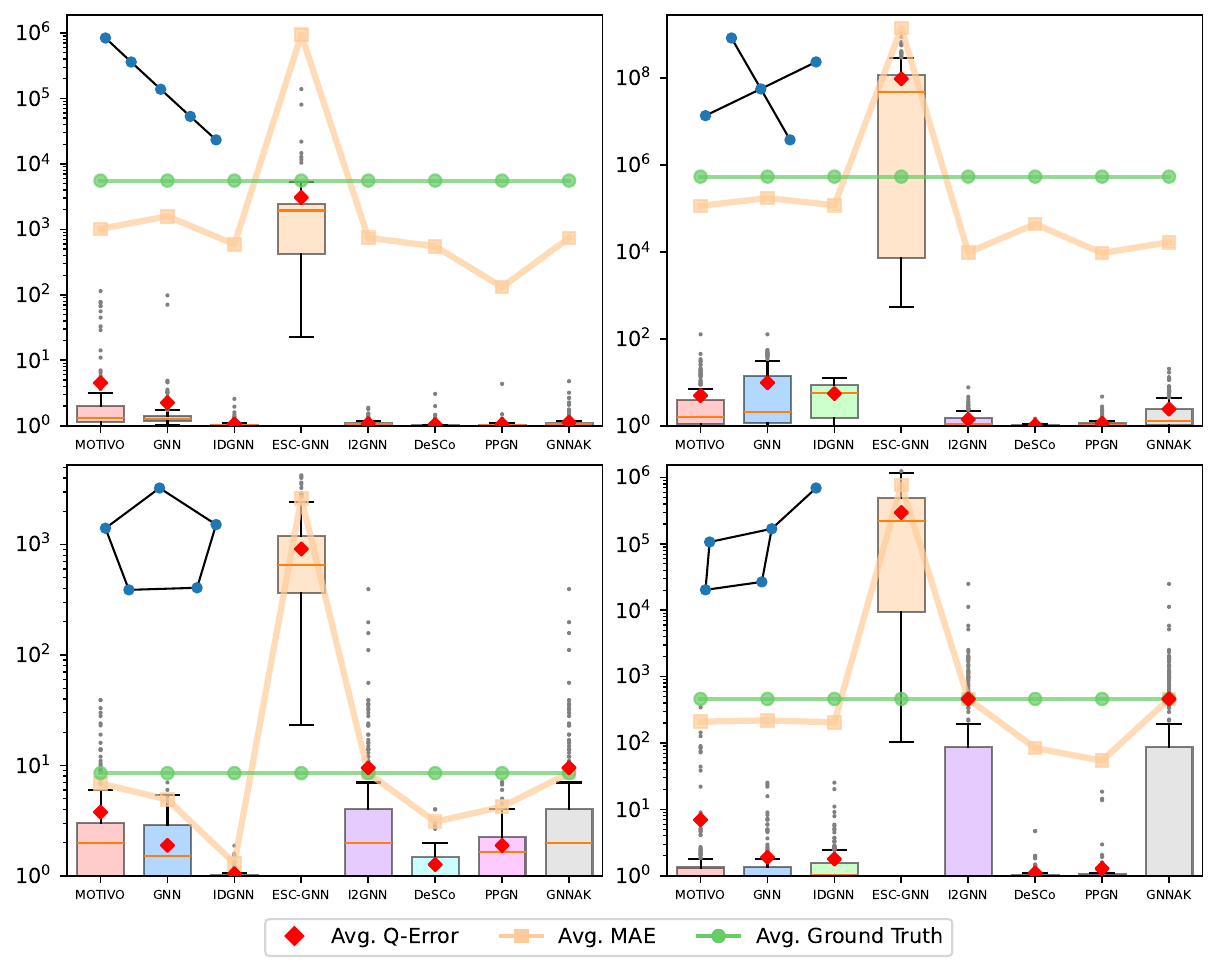}
    }
  \caption{Accuracy on different training scenarios.}
  \vspace{-0.17in}
  \label{fig:accuracy_precision}
\end{figure*} 

\subsection{Results}
In this section, we present the experimental results of our benchmark. Our results are evaluated using several performance metrics, including efficiency, accuracy, and robustness. We begin by assessing the runtime performance of each method to measure their computational efficiency. Next, we evaluate how accurately the models estimate subgraph counts across various graph types. Finally, we examine the models' robustness by testing them under different graph conditions, which helps us understand their ability to generalize to unseen patterns and datasets. Each part of the evaluation provides key insights into the strengths and limitations of the methods.

\begin{figure}[h]
      \centering
      \includegraphics[width=\columnwidth]{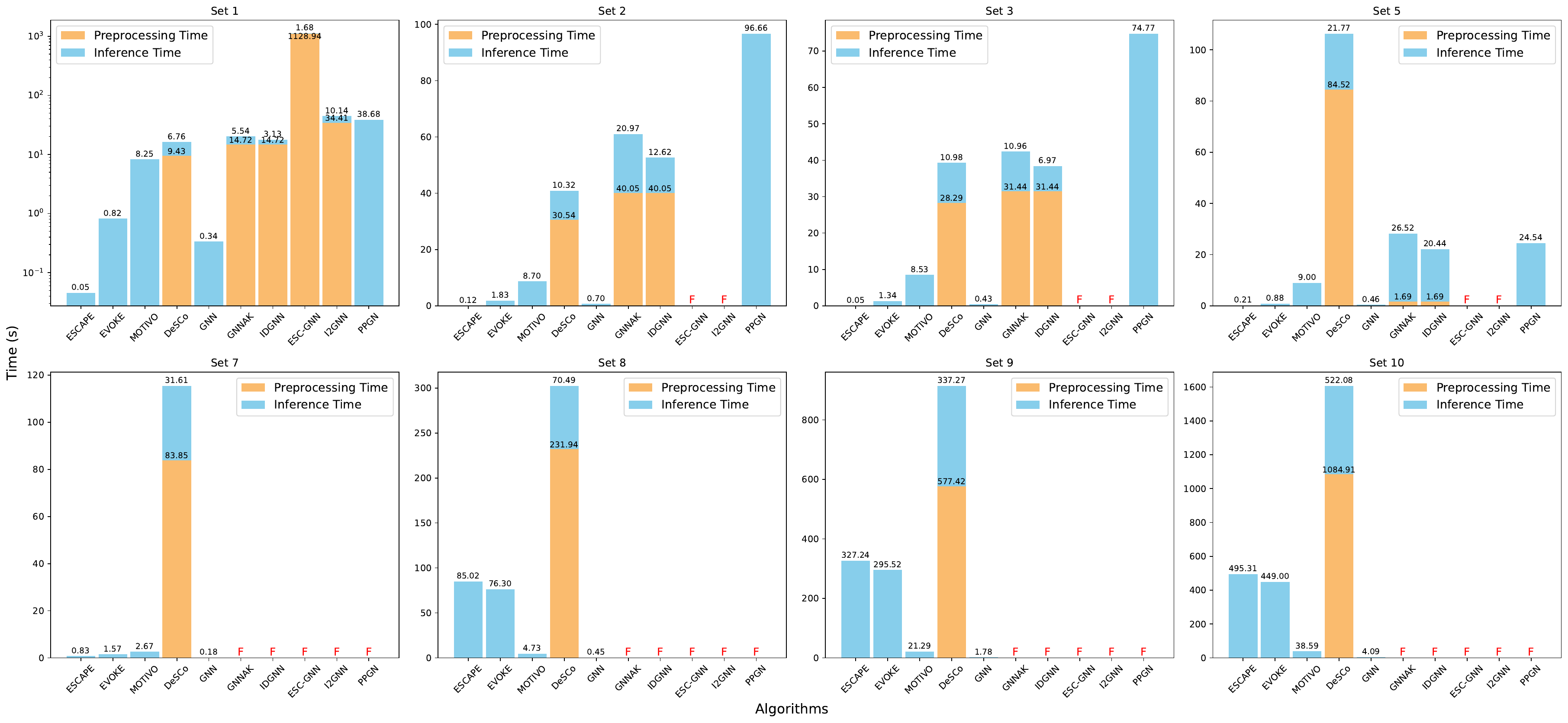}
      \vspace{-0.2in}
      \caption{Inference and pre-processing time.}
      \vspace{-0.17in}
      \label{fig:runtime}
\end{figure}

\textbf{Efficiency}
We evaluate the efficiency of our algorithms across various datasets and pattern complexities. The overall runtime is broken down into three phases: \textbf{pre-processing}, \textbf{training}, and \textbf{inference}. Pre-processing and inference times are highlighted in Figure \ref{fig:runtime}, while Table \ref{tab:trainingtime} shows a detailed comparison of the training times for models that require it. 

This separation is important because, in practical scenarios, a well-trained model should be able to generalize to new graphs without any additional fine-tuning—meaning only pre-processing and inference would be needed. If additional training is required, it indicates that ground truth data must be accessible, which would defeat the purpose of using an ML model for prediction in the first place.

\begin{table}[ht]
    \centering
    \small
    \begin{tabular}{|@{~}c@{~}|@{~}c@{~}|@{~}c @{~}|@{~}c@{~}|@{~}c@{~} |@{~}c@{~} |@{~}c@{~} |@{~}c @{~}|@{~}c@{~}|}
        \hline
        \textbf{Algorithms} & \textbf{Set\_1} & \textbf{Set\_2} & \textbf{Set\_3}&\textbf{Set\_5}&\textbf{Set\_7}&\textbf{Set\_8}&\textbf{Set\_9}&\textbf{Set\_10} \\
\hline
\textbf{DeSCo-ST}&269& 500 &616& 1705& 734 &2069 &3578 &12167\\
\hline
\textbf{DeSCo}&22.7& 37.8 &47.6 &102 &44.2& 110& 211 &639\\
\hline
\textbf{GNN}&2.37& 4.52& 2.97 &3.24& 1.06& 1.86& 3.59& 9.66\\
\hline
\textbf{GNNAK}&81.4 &255 &145 &336 &N/A &N/A &N/A &N/A\\
\hline
\textbf{IDGNN}&40.7& 145 &86.8& 238& N/A &N/A &N/A &N/A\\
\hline
\textbf{ESC-GNN}&33.5& N/A& N/A &N/A &N/A& N/A& N/A& N/A\\
\hline
\textbf{I2GNN}&118 &N/A &N/A &N/A &N/A &N/A &N/A &N/A\\
\hline
\textbf{PPGN}&636 &1410& 1197& 446 &N/A &N/A &N/A& N/A\\
\hline
    \end{tabular}
    \caption{Training time comparison (minutes).}
    \label{tab:trainingtime}
\end{table}

\paragraph{Training Time (Table~\ref{tab:trainingtime})}

Table~\ref{tab:trainingtime} presents a comparison of training times for the different methods across datasets of various sizes. Here, "N/A" indicates that a method failed during training, primarily due to GPU memory constraints. The results reveal several key trends:

PPGN consistently exhibits the highest training times across all datasets, with its training duration increasing significantly as the dataset size grows. Despite these longer training times, PPGN remains one of the most accurate and robust architectures for subgraph counting. It is important to emphasize that the reported training times are averaged over all 29 subgraphs (targets), meaning that the values indicate the average time (in minutes) required to train the model for a single target.

One notable trend is that the vanilla GNN exhibits the lowest training time among all methods, as shown in Table~\ref{tab:trainingtime}. This efficiency comes from its simple architecture, making it the most scalable model as graph sizes increase. Such scalability offers promise for addressing the challenges that traditional AL methods face with large graphs. However, despite its low training time, the accuracy of the vanilla GNN is lower compared to more complex models. This is mainly because its discriminative power is limited to 1-WL \cite{DBLP:conf/iclr/XuHLJ19}.

DeSCo is originally designed to train on all targets concurrently. To facilitate a fair comparison, we modified DeSCo to focus on a specific target by training sequentially on each target and summing the training times. The corresponding training time for a single target under this configuration is noted as DeSCo-ST (Separate Target). As expected, the vanilla message-passing GNN remains the simplest and fastest model to train.

Another critical observation is that many of these architectures demand significant memory resources, with some methods running out of GPU memory on larger graphs even when using batch sizes as small as \(1\). Additionally, with all models except DeSCo, training on Set\_2 consistently takes longer than on Set\_3. This performance difference is due to the higher density of graphs in Set\_2, which directly impacts the training time. Overall, these observations underscore the inherent trade-offs between training efficiency, memory usage, and the level of architectural complexity in subgraph counting tasks.

\paragraph{Pre-processing and Inference Time (Figure~\ref{fig:runtime})}
Figure \ref{fig:runtime} shows the \textbf{pre-processing} and \textbf{inference} times for the evaluated algorithms on different datasets. In practical applications, where models are already well-trained and do not require additional fine-tuning, only the steps of preparing the data (pre-processing) and making predictions (inference) are needed.

Algorithms like ESCAPE, EVOKE, GNN, and MOTIVO either avoid pre-processing or need very little of it. This figure highlights several important points:

First, pre-processing time must be considered as part of the overall runtime for subgraph counting. Ignoring this step can lead to an unfair comparison, especially if a method performs significant subgraph counting during the pre-processing phase.

Second, an efficient architecture should aim to minimize pre-processing time—an area where many current architectures still have room for improvement.

Third, the vanilla GNN shows a very promising running time, particularly on large graphs. Its sublinear growth in running time suggests that GNN models could be developed further to handle larger graphs efficiently, especially if the current challenges with accuracy (related to its limited discriminative power with the 1-WL framework) can be overcome.

\textbf{Accuracy and Precision}
We evaluate the accuracy of existing algorithms across three different scenarios, each designed to highlight a specific aspect of the method's performance. Figure~\ref{fig:accuracy_precision} presents boxplots comparing the performance of the methods under these scenarios. The evaluation metrics include the average Q-error, average MAE, and average ground truth. It is important to note that MOTIVO, a non-learning approximate counting algorithm, remains unaffected by these scenarios. Rather, it serves as a baseline to indicate the accuracy scale and facilitate comparisons with the learning-based methods. It is important to note that Set\_1 is the dataset used for this figure since all the methods can support it.

In the plot, the boxes represent the Q-error, while the red dots denote the average Q-error. The orange line indicates the average MAE, and for better context regarding the scale of the MAE, we also include the average ground truth (green line).

\textbf{Zero-Shot Test [Fig~\ref{fig:accuracy_precision}a]:} We pretrained all models on the dataset provided in ESC-GNN \cite{esc} using the suggested hyperparameters for each method. This experiment reveals the generalizability of the algorithms, determining whether they can truly ``learn to count'' or if they merely learn the distribution of counts from the training set. This shows that the learning ability of the existing algorithms varies significantly among different patterns.

\textbf{Few-Shot Fine-Tuning [Fig~\ref{fig:accuracy_precision}b]:} In this test, we take the pretrained models from the previous experiment and fine-tune them using a training set similar to the test set. We fine-tune the model for a few shots (10\% of the full training data), then evaluate its accuracy. This scenario reveals how quickly the models can adapt to new datasets with limited training examples and assesses how well they have learned to count.

\textbf{Few-Shot Retrain [Fig~\ref{fig:accuracy_precision}c]:} This test is similar to the previous one, but instead of using a pretrained model, we train the models from scratch. The goal is to compare whether the model learns better when fine-tuned or when trained from scratch with few training examples.

\begin{table}[h]
    \centering
    \small
    \begin{tabular}{|c|c|c|c|c|}
        \hline
        \textbf{Algorithms} & \textbf{Set\_1} & \textbf{Set\_2} & \textbf{Set\_3} & \textbf{Set\_5} \\
        \hline
        \textbf{Motivo}   & 108.88  & 469.86   & 2.43   & 1.13   \\
        \hline
        \textbf{DeSCo}    & 1.93    & 1.39     & 1.13   & 1.21   \\
        \hline
        \textbf{GNN}      & 8.33    & 24.63    & 7.69   & 2.23   \\
        \hline
        \textbf{GNNAK}    & 4307.59 & 7612.56  & 173.02 & 1438.93\\
        \hline
        \textbf{IDGNN}    & 4.45    & 26.48    & 1.12   & 1.76   \\
        \hline
        \textbf{ESC-GNN}  & 29220.07 & N/A      & N/A    & N/A    \\
        \hline
        \textbf{I2GNN}    & 5.01    & N/A      & N/A    & N/A    \\
        \hline
        \textbf{PPGN}     & 1.02    & 4.01     & 1.39   & 1.64   \\
        \hline
    \end{tabular}
    \caption{Few-Shot Q-Error on target 13.}
    \label{tab:fewshot_qerror}
\end{table}

\begin{figure}[h]
      \centering
      \includegraphics[width=0.85\columnwidth]{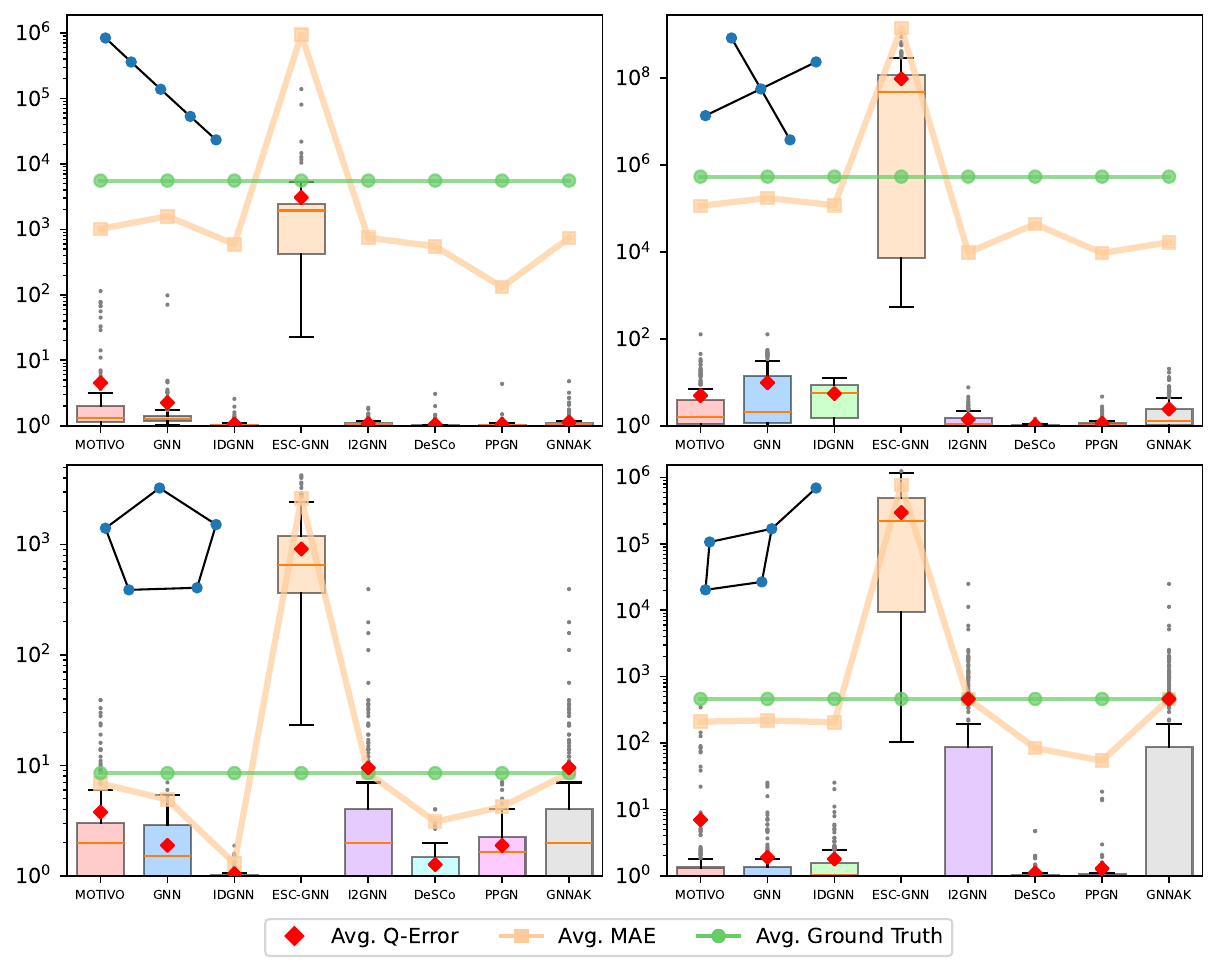}
      \vspace{-0.2in}
      \caption{Full Training Accuracy Comparison.}
      \vspace{-0.17in}
      \label{fig:full_retrain_acc}
\end{figure}

\textbf{Full Training [Fig~\ref{fig:full_retrain_acc}]:} This test evaluates the performance of each algorithm in an ideal training scenario, where the models are fully trained from scratch using the entire dataset. This experiment serves as a baseline and is typically the approach used in most papers to claim superiority of one model over others.

Since in practice it doesn’t make sense to train the model before counting (as it is very time-consuming), and if we already know the ground truth (GT) for training, we wouldn't need the prediction model. Additionally, since the zero-shot accuracy of the models is not very promising, and few-shot fine-tuned models generally perform better than few-shot retrained ones on average (as shown in Figure~\ref{fig:accuracy_precision}), we chose fine-tuning as the optimal solution.

We fine-tuned all of our models and evaluated their accuracy using the Q-error metric, as presented in Table~\ref{tab:fewshot_qerror}. In the table, "N/A" indicates that the model ran out of memory while processing that dataset. As can be seen, the overall winners in terms of accuracy are PPGN and DeSCo.

\textbf{Robustness and Generalization}
In this section, we evaluate the robustness and generalization capabilities of the models under various conditions. Robustness is crucial for ensuring that the model performs reliably even when faced with very dense nodes or graphs, while generalization assesses the model's ability to adapt to unseen graph structures. We analyze the model's performance on our benchmark datasets, focusing on its response to varying graph properties such as degree, density, and node clustering coefficients. 

\paragraph{Graph Structure Impact on Accuracy} 
We begin by examining how different graph structural properties, including degree and density, influence the model's accuracy.

\begin{figure}[ht]
  \centering
  \subfigure[Graph]{
   \label{fig:graph_heat}
    \includegraphics[width=0.45\textwidth]{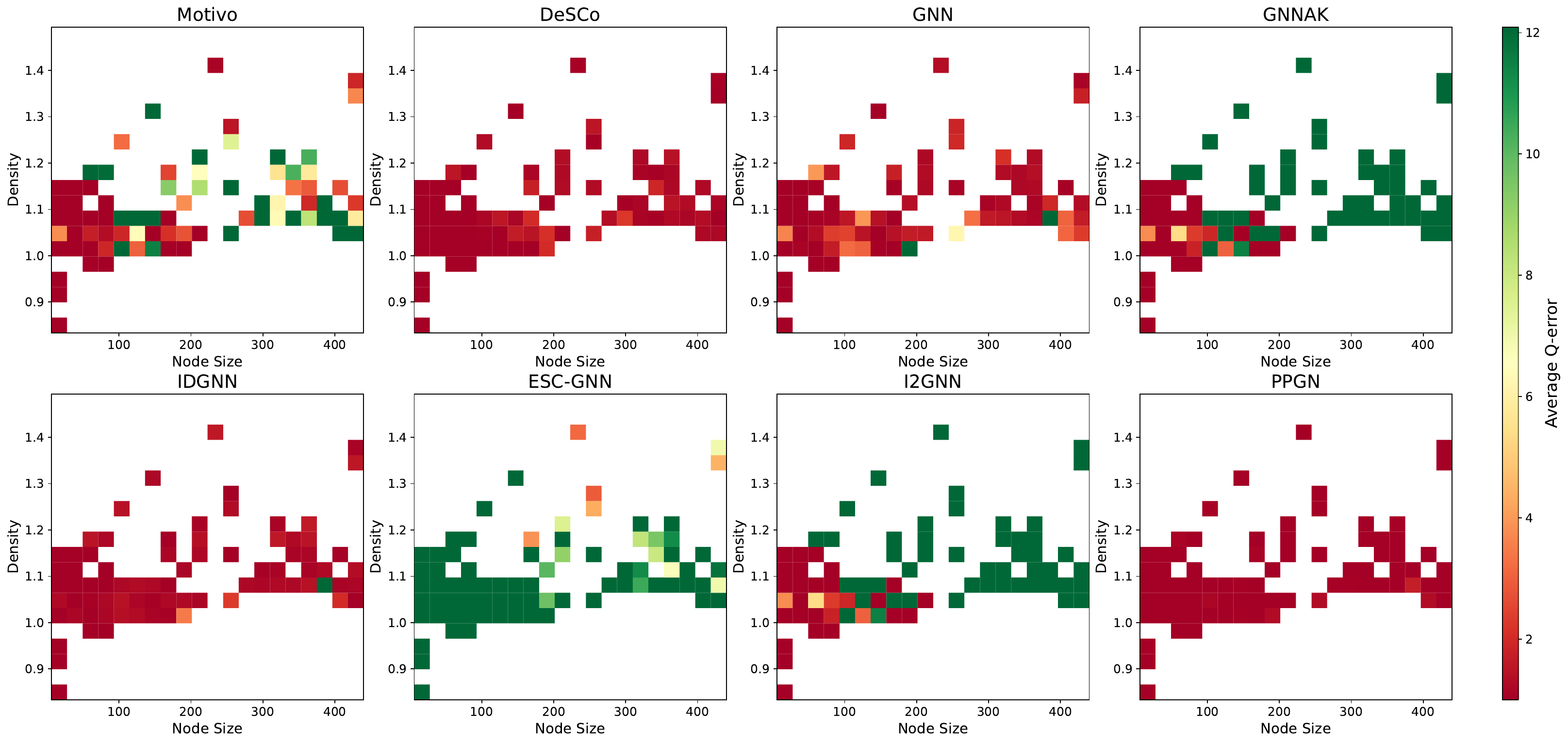}
    }\!\!
  \subfigure[Node]{
   \label{fig:node_heat}
    \includegraphics[width=0.45\textwidth]{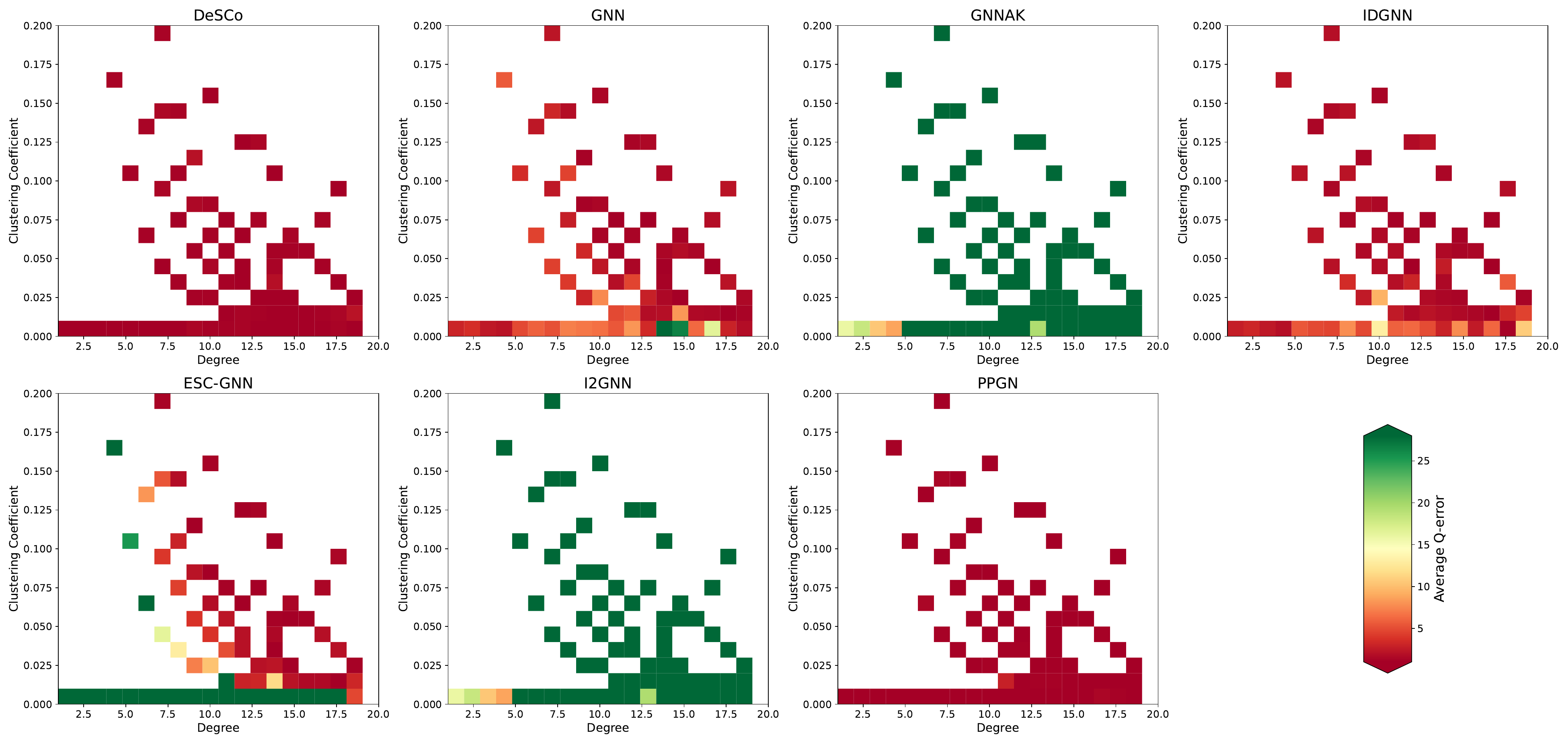}
    }
    \vspace{-0.2in}
  \caption{Effect of graph size and density and node degree and clustering coefficient on accuracy.}
  \vspace{-0.17in}
  \label{fig:graph_node_heat}
\end{figure}

Figure \ref{fig:graph_node_heat} illustrates the impact of graph-level (Subfigure \ref{fig:graph_heat}) and node-level (Subfigure \ref{fig:node_heat}) characteristics on subgraph counting accuracy. In Subfigure \ref{fig:graph_heat}, each point represents a graph, with the x-axis indicating the graph size (in terms of the number of nodes, |V|) and the y-axis reporting the graph density ($\frac{|E|}{|V|}$). The color of each point corresponds to the Q-Error (with lower values being better). This plot demonstrates how each method reacts to global graph features. The underlying assumption is that larger graphs with higher density tend to make subgraph count prediction more challenging—a trend that holds for most methods, although there are notable exceptions (e.g., in some regions for GNN or ESC-GNN).

Similarly, Subfigure \ref{fig:node_heat} focuses on local graph properties. Here, each point represents a node, the x-axis shows the node degree, and the y-axis corresponds to the node's clustering coefficient. The color again reflects the Q-Error. This subfigure is used to analyze how local node features influence prediction accuracy, with the general assumption that nodes with higher degrees and clustering coefficients will yield less accurate counts. While this pattern generally holds true, some methods, such as ESC-GNN, occasionally deviate from this expectation.

\paragraph{Learning Efficiency} 
Next, we evaluate the learning efficiency of the models by analyzing how quickly and accurately it converges during training. The convergence speed is crucial for assessing the model's robustness over time, as well as its stability during long-term training.

\begin{figure}[h]
      \centering
      \includegraphics[width=\columnwidth]{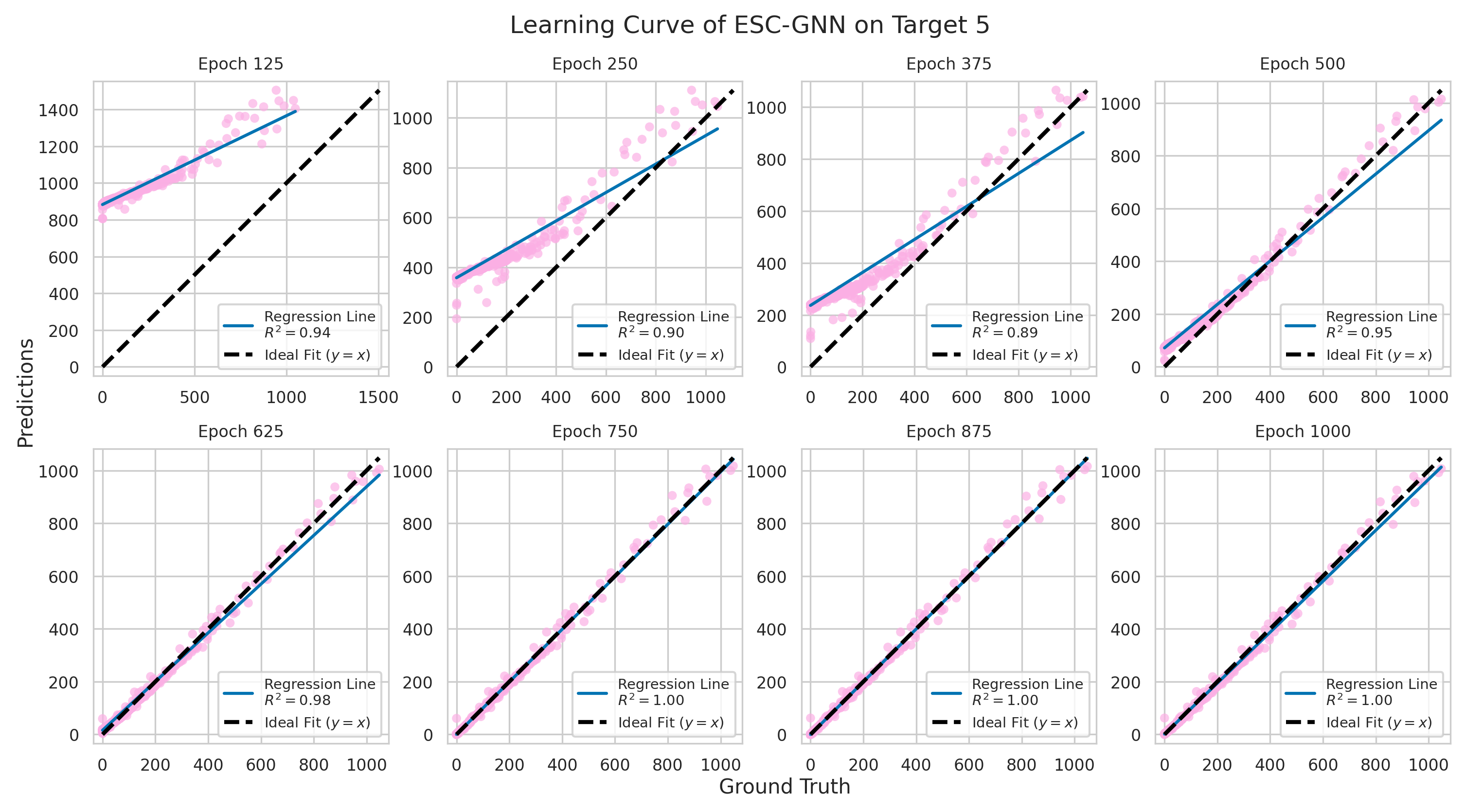}
      \vspace{-0.2in}
      \caption{Learning curve showing how fast and accurately ESC-GNN converges.}
      \vspace{-0.17in}
      \label{fig:learning_curve}
\end{figure}

Figure \ref{fig:learning_curve} shows the learning curve for ESC-GNN, demonstrating how it converges and stabilizes after approximately 1000 epochs. While ESC-GNN requires a longer training period for stability, its accuracy improves significantly over time. This highlights the importance of extended training for achieving optimal performance. For a complete comparison of the learning curves of other algorithms, please refer to our website.

\paragraph{Finetuning and Adaptability} 
We further assess the ability of different models to adapt and finetune to specific tasks, focusing on their performance on a particular target (target 11).

\begin{figure}[h]
  \centering
  \subfigure[Final results]{
   \label{fig:finetune_results}
    \includegraphics[width=0.43\textwidth]{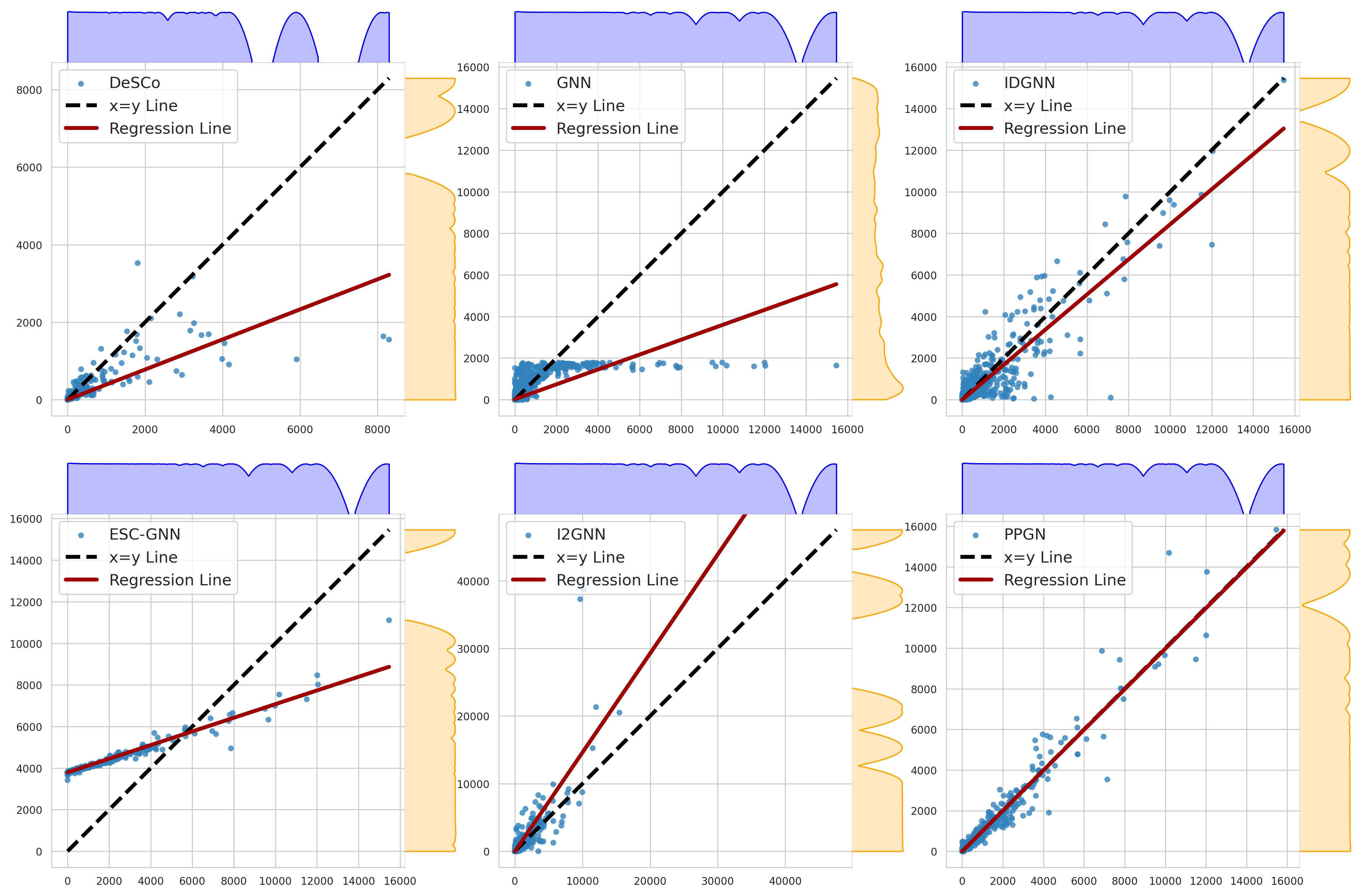}
    }\!\!
  \subfigure[Q-Error trend]{
   \label{fig:qerror_trend}
    \includegraphics[width=0.47\textwidth]{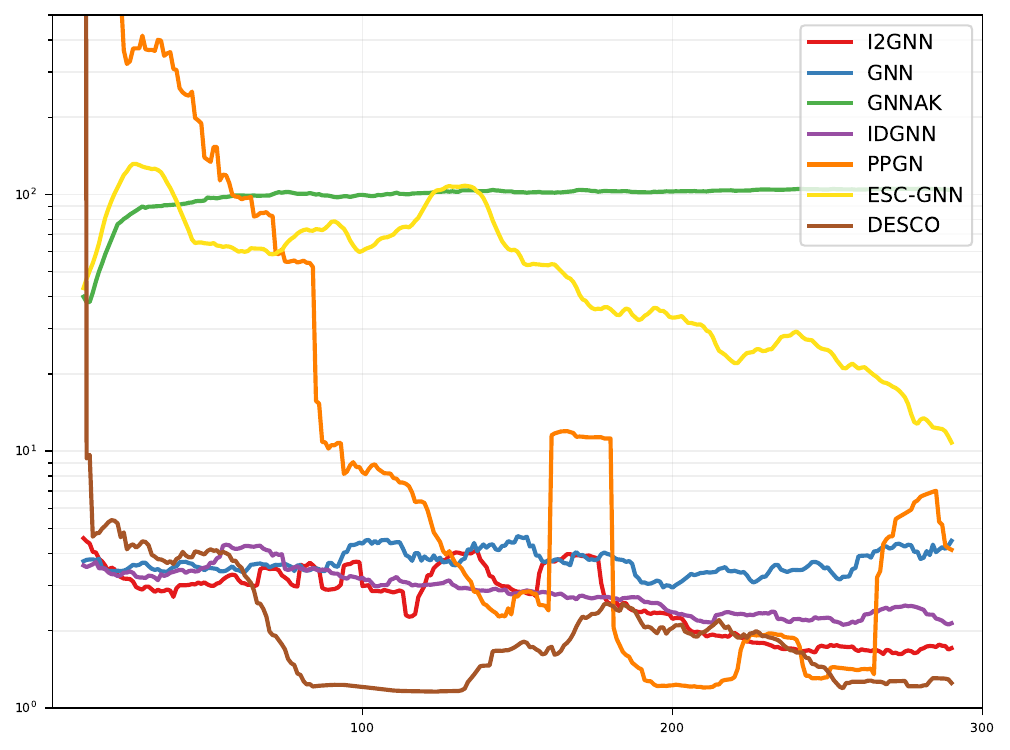}
    }
    \vspace{-0.2in}
  \caption{Finetuning ability and Q-Error trend on Set 1, target 11.}
  \vspace{-0.1in}
  \label{fig:finetuning_ability}
\end{figure}

Figures \ref{fig:finetune_results} and \ref{fig:qerror_trend} present the finetuning results and Q-Error trend for different models on target 11. The scatter plot in Figure \ref{fig:finetune_results} shows that certain models, such as ESC-GNN and I2GNN, demonstrate superior finetuning capabilities, closely aligning their predictions with the ground truth. The Q-Error trend in Figure \ref{fig:qerror_trend} further highlights the reduction in Q-Error over time, indicating that models with better adaptability improve their accuracy through finetuning. These results emphasize the model's generalization ability and its capacity to adjust to task-specific requirements.

\paragraph{Accuracy vs. Pattern Size} 
We compare the performance of different models on varying subgraph patterns, focusing on 3-star, 4-star, and 5-star patterns. Subgraph pattern recognition is a key indicator of a model's ability to generalize across different graph structures.

\begin{table}[h]
    \centering
    \scriptsize
    \begin{tabular}{|c|c|c|c|c|c|c|}
        \hline
        \multicolumn{1}{|c|}{\textbf{Alg}} & \multicolumn{2}{c|}{\textbf{3-star}} & \multicolumn{2}{c|}{\textbf{4-star}} & \multicolumn{2}{c|}{\textbf{5-star}} \\
        \cline{2-7}
        & \textbf{MAE} & \textbf{Q-Error} & \textbf{MAE} & \textbf{Q-Error} & \textbf{MAE} & \textbf{Q-Error} \\
        \hline
        \textbf{Motivo}   & $1.1e+02$ & $1.23$ & $4.0e+03$ & $1.72$ & $1.1e+05$ & $4.97$ \\
        \hline
        \textbf{DeSCo}    & $1.7e+02$ & $1.07$ & $2.9e+03$ & $1.07$ & $4.4e+04$ & $1.25$ \\
        \hline
        \textbf{GNN}      & $7.1e+02$ & $1.84$ & $1.0e+04$ & $1.64$ & $1.9e+05$ & $10.00$ \\
        \hline
        \textbf{GNNAK}    & $7.4e+02$ & $1.93$ & $5.7e+03$ & $1.24$ & $2.4e+04$ & $2.46$ \\
        \hline
        \textbf{IDGNN}    & $5.1e+02$ & $1.64$ & $6.7e+03$ & $1.25$ & $1.2e+05$ & $2.47$ \\
        \hline
        \textbf{ESC-GNN}  & $1.3e+06$ & $4.3e+03$ & $4.9e+06$ & $3.9e+04$ & $3.0e+07$ & $2.1e+06$ \\
        \hline
        \textbf{I2GNN}    & $5.2e+02$ & $1.65$ & $6.0e+03$ & $1.25$ & $5.1e+03$ & $2.42$ \\
        \hline
        \textbf{PPGN}     & $5.2e+02$ & $1.65$ & $5.1e+03$ & $1.20$ & $2.8e+04$ & $2.18$ \\
        \hline
    \end{tabular}
    \caption{Star pattern accuracy with MAE and Q-Error for 3-star, 4-star, and 5-star patterns.}
    \label{tab:starpattern}
\end{table}

Table \ref{tab:starpattern} presents the accuracy (in terms of MAE and Q-Error) for 3-star, 4-star, and 5-star patterns across various models. We observe that models like DeSCo and I2GNN consistently perform well across all patterns, whereas others like ESC-GNN exhibit significant deviations, particularly with larger patterns. This analysis highlights potential accuracy variations on patterns of similar nature with different sizes, with some models favoring accuracy for smaller patterns while others generalize better to larger patterns.

\section{Conclusions} \label{sec:concl}
In this paper, we proposed an efficient and scalable benchmark called BEACON,
a scalable and standardized benchmark designed to evaluate tradi-
tional and machine learning-based subgraph counting algorithms.
BEACON tackles key issues such as
reproducibility, scalability, and generalizability. Through extensive
experiments, we evaluated the utility of BEACON in identifying the
strengths and weaknesses of various algorithms, emphasizing the
need for a universal benchmarking framework. In the future, we plan to investigate 
the limitations and open challenges of traditional algorithms as
well as current machine learning-based techniques for subgraph
counting, highlighting opportunities for future research.

\clearpage

\bibliographystyle{ACM-Reference-Format}
\bibliography{bib/references, bib/datasets}

\end{document}